\documentclass[12pt]{article}
\usepackage{amsmath,amssymb,epsfig}   

\begin{document}

\title {Design studies for a European Gamma-ray Observatory}  
\maketitle  

 Carmen Baixeras$^b$,
 Denis Bastieri$^h$,
 Wlodek Bednarek$^e$,
 Ciro Bigongiari$^h$,
 Adrian Biland$^p$,
 Oscar Blanch$^a$,
 Rudolf K. Bock$^g$,
 Thomas Bretz$^m$,
 Ashot Chilingarian$^n$,
 Jos\'{e} Antonio Coarasa$^g$,
 Sebastian Commichau$^p$,
 Luis Jos\'{e} Contreras$^f$,
 Juan Cortina$^a$,
 Francesco Dazzi$^h$,
 Alessandro De Angelis$^o$,
 Barbara De Lotto$^o$,
 Carles Domingo$^b$,
 Eva Domingo$^a$,
 Daniela Dorner$^m$,
 Daniel Ferenc$^d$,
 Enrique Fern\'{a}ndez$^a$,
 Josep Flix$^a$,
 Victoria Fonseca$^f$,
 Lluis Font$^b$,
 Nicola Galante$^k$,
 Markus Gaug$^a$,
 J\"urgen Gebauer$^g$,
 Riccardo Giannitrapani$^o$,
 Maria Giller$^e$,
 Florian Goebel$^g$,
 Thomas Hengstebeck$^j$,
 Piotr Jacon$^e$,
 Okkie C. de Jager$^i$,
 Oleg Kalekin$^c$,
 Daniel Kranich$^d$,
 Elina Lindfors$^l$,
 Francesco Longo$^o$,
 Marcos L\'{o}pez$^f$,
 Javier L\'{o}pez$^a$,
 Eckart Lorenz$^g$,
 Fabrizio Lucarelli$^f$,
 Karl Mannheim$^m$,
 Mos\`{e} Mariotti$^h$,
 Manel Mart\'{i}nez$^a$,
 Keiichi Mase$^g$,
 Martin Merck$^m$,
 Mario Meucci$^k$,
 Razmick Mirzoyan$^g$,
 Abelardo Moralejo$^h$,
 Emma O\~{n}a-Wilhelmi$^f$,
 Raul Ordu\~{n}a$^b$,
 David Paneque$^g$,
 Riccardo Paoletti$^k$,
 Mikko Pasanen$^l$,
 Donatella Pascoli$^h$,
 Felicitas Pauss$^p$,
 Nikolaj Pavel$^j$,
 Raffaelo Pegna$^k$,
 Luigi Peruzzo$^h$,
 Alessio Piccioli$^k$,
 Massimo Pin$^o$,
 Raquel de los Reyes$^f$,
 Arnau Robert$^b$,
 Antonio Saggion$^h$,
 Alejandro S\'{a}nchez$^b$,
 Paolo Sartori$^h$,
 Villi Scalzotto$^h$,
 Aimo Sillanp\"{a}\"{a}$^l$,
 Dorota Sobczynska$^e$,
 Antonio Stamerra$^k$,
 Leo Takalo$^l$,
 Masahiro Teshima$^g$,
 Nadia Tonello$^g$,
 Andreu Torres$^b$,
 Nicola Turini$^k$,
 Gert Viertel$^p$,
 Vincenzo Vitale$^g$,
 Serguei Volkov$^j$,
 Robert Wagner$^g$,
 Tadeusz Wibig$^e$,
 Wolfgang Wittek$^g$    \\ 

\par (a) {Institut de Física d'Altes Energies, Barcelona} 
\par (b) {Universitat Autonoma de Barcelona} 
\par (c) {Crimean Astrophysical Observatory}
\par (d) {University of California, Davis}
\par (e) {Division of Experimental Physics, University of Lodz}
\par (f) {Universidad Complutense, Madrid}
\par (g) {Max-Planck-Institut f\"ur Physik, M\"unchen}
\par (h) {Dipartimento di Fisica, Universit\`{a} di Padova}
\par (i) {Space Research Unit, Potchefstroom University}
\par (j) {Fachbereich Physik, Universit\"at-GH Siegen}
\par (k) {Dipartimento di Fisica, Universit\`{a} di Siena}
\par (l) {Tuorla Observatory, Pikki\"o}
\par (o) {Dipartimento di Fisica, Universit\`{a} di Udine}
\par (m) {Universit\"at W\"urzburg} 
\par (n) {Yerevan Physics Institute, Cosmic Ray Division, Yerevan}
\par (p) {Institute for Particle Physics, ETH Z\"urich}

\section*{Abstract}  

In this note we discuss preliminary studies concerning 
a large-diameter gamma-ray telescope, to be part of an array of telescopes 
installed at the existing observation site on the Canary island of La Palma. 
One of the telescopes in the array
will be MAGIC, presently the largest existing gamma ray
telescope with the lowest energy threshold world wide.
A second telescope of the same class
is under construction. Eventually, we will want to install one or more 
devices giving access to even lower gamma-ray energy; they will be 
larger than MAGIC by roughly a linear factor two, and are code-named 
ECO-1000 (for a mirror surface of 1000~m$^2$). 
\par
A lower energy threshold is the key to new understanding in the 
gamma-ray domain observable by
ground-based Cherenkov telescopes. It will
allow to cover wavelengths in overlap with foreseen (and past)
satellite experiments. We discuss below the substantial physics 
potential made available by a lower energy threshold. 
We also show how larger telescopes and higher light collection efficiency
can lower the observable energy threshold, substantiated by extensive simulations.
The simulations also confirm that multiple telescopes, of the MAGIC or the 
larger class, can achieve higher sensitivity. 
\par
We discuss the technologies needed to reach the physical low-energy limit. 
They exist
at the component level, but have to be field-tested; we 
propose to implement and integrate the most critical components 
in a MAGIC-class telescope, such that the
eventual extrapolation to a larger device becomes a fully predictable step. 
If financing can be found, such 
tests can be completed on a timescale such that a proposal for the
first true low-theshold ($\leq$8~GeV) telescope can be made in 2007, 
and its construction completed in 2009/2010.
Together, the future telescopes will constitute a European 
Cherenkov Observatory (ECO) with unprecedented attraction for the worldwide
research community, a step we deem natural after 
having successfully installed MAGIC in 2003.

\section{Executive Summary}

We discuss a feasibility study concerning several
critical prototype elements for a large telescope. We think of a telescope with a 
useful mirror surface of about 1000~m$^2$, called ECO-1000, which is 
characterized, compared to the existing MAGIC telescope, by 

\begin{itemize} 
  \item     improved and wavelength-extended sensitivity of photomultipliers
  \item     fourfold mirror surface achieved with new technologies
  \item     permanently active mirror focusing under computer control
  \item     support structure of low weight, allowing accelerated telescope movement
  \item     economic high-performance readout electronics, including signal processing
\end{itemize}
\par 

In more detail, we plan detailed studies and prototyping work in several 
areas:

\begin{itemize}
\item  Light collection: 
large-surface photomultiplier tubes with high quantum efficiency and 
wavelength acceptance extending into the UV;
dielectric foils for optimal light collection (Winston cones);
 new production methods for hexagonal mirror elements with an individual surface in
excess of one square meter; mirror surfaces with the highest possible reflectivity;
 automated and robust methods for focusing the mirror elements

 \item  Telescope structure: 
  new materials for building a light-weight support structure;
  accelerated telescope drive motors;
  simulation of structure deformation, wind resistance and possible oscillations;
  wind protection of structure (e.g. a low-resistance cover and/or light-weight 
  clam shell mirror-only dome, similar to that of the Liverpool 2m telescope on La Palma); 
  anchoring of the azimuth rail with minimal deformation.

\vspace{1mm} \item   Site related studies: 
  geological and meteorological suitability of possible sites near MAGIC;
  estimation of additional needs for infrastructure like electrical power, 
  roads, control room surface, lightning protection; environmental impact.

 \item   Dissemination and data access studies: 
we believe that a multi-telescope observatory must provide access to the data to a 
community much wider than the collaboration that builds the detectors. At a 
time when it becomes obvious that much progress is to be made by multi-wavelength
observations, this seems an obvious necessity. We want to adapt our operation 
structure and format, access and dissemination of our data to a broad international
scientific community.

 \item  Electronics and data acquisition: 
  more economical and less power-hungry multi-channel readout with 
  sub-nanosecond time resolution 
  and full preservation of digital signal properties, with real-time analysis 
  in programmable processing units;
  adaptation of the data interface to modern computer communication 
(e.g. Gigabit Ethernet or firewire): 
this is an area of extraordinarily rapid evolution and potential 
cost cutting;

 \item Physics: 
  modeling of different physics processes;
  precise estimation of achievable energy and angular resolution, 
  and their impact on the various physics goals;
  performance comparison of an ECO-1000 telescope with foreseen satellite experiments.
\end{itemize}

Many of these studies are guided by our experience in building and commissioning
MAGIC. Not all of them are technologically challenging as to require
more than the developments we will carry out in the framework of building a copy of
MAGIC, itself already containing multiple innovations.
\par
The experience of developing MAGIC
and bringing into operation this telescope, which has the presently lowest 
energy threshold of all gamma ray telescopes, will be invaluable for the next steps.
The European La Palma site is already home for MAGIC, and has been so, 
in the past, for several
gamma ray and air shower experiments. The site is also being used by 
multiple first-rate optical telescopes. The location has 
clear advantages over many other sites, including those proposed at high altitude. 
With MAGIC and the proposed telescopes, La Palma will become 
an observatory for gamma rays, unequalled in the Northern Hemisphere. 
We are convinced that the development of ECO will make 
major contributions in resolving present questions
in fundamental physics and cosmology. 
\par
We explicitely want to express our openness to a close scientific
collaboration with other experiments (HESS, Veritas, Cangaroo),
following a development line close to that of MAGIC,
with a view of installing optimal gamma ray observation possibilities 
in the Northern and Southern Hemisphere. 
We would expect the eventual instruments to run 
in close collaboration.
\par
The total cost of the studies should be of the order of
5~MEuro, the timescale expected is not more than 3~years. 
A detailed proposal for a large-size device (ECO-1000) 
will be completed toward the end of this period, viz. in 2007 or 2008, 
and construction can start shortly afterwards,
for a possible completion in 2009/2010. The cost  of ECO-1000 
is estimated today at 15-20~MEuro. 

\section{Introduction} 

MAGIC was designed back in 1998 with the very
clear goal to lower the energy threshold at which gamma rays can be observed
\cite{MagicDR}. 
As the first MAGIC telescope has started operation and is now, in early 2004,
on the way to reach its design performance, 
it is likely that, beyond the physics that has been predicted, 
unexpected avenues will open. Many unanswered physics questions loom in the
low energy range, and we are confident that several of them will already be 
answered by MAGIC.  The instruments of ECO, in particular the eventual larger 
telescope(s), will subsequently be able to extend the energy limit to even 
lower values, close to the lowest energies accessible to the Cherenkov technique,
and to fully explore this range of wavelengths. 

The energy threshold is primarily a question of two components: 
photon detection efficiency and mirror surface. Fast electronics resulting in signal 
integration over a short time is also an 
important handle to suppress background, and contributes to observing
showers at lower energies.
A natural evolution of the MAGIC principles, therefore, will be to take advantage of 
technological progress in light detection and the building of large structures, viz. the
construction of a larger and more performant telescope. The 
mirror surface must be pushed to the technical limit without taking undue risks. To be 
constructed with a large mirror surface, maintaining a fast slewing possibility,
the telescope structure must be made light-weight and easy to assemble. 
For high sensitivity, photodetectors must be brought
to the highest possible quantum efficiency, and the light collection along the 
optical path must be optimal.

These are goals that have already guided the construction of MAGIC. 
The addition to MAGIC of further telescopes, on the
Roque de los Muchachos site (La Palma), 
constitutes the initial step in building up a gamma ray observatory,  
the European Cherenkov Observatory or ECO.
The experience gained during the construction of MAGIC, and progress in industrially
available components and production methods do allow an improvement 
program, for a future second telescope of the MAGIC type. 
A so far rather unchanged copy of MAGIC is already under way. 

With some additional support, this telescope can incorporate 
key innovations in technology, 
viz. the latest commercially available photodetectors, a new light-weight 
structure supporting highly reflective aspherical mirrors, and 
much improved mirror control that does not interfere with observations. 
Prototypes developed in that framework, and an accompanying 
feasibility and design study will pave the way for 
a substantial future leap in telescope size, i.e. towards ECO-1000. 

First results from the MAGIC observatory are being obtained now (early 2004),
and will likely turn out to be the start of a bonanza 
of sources understood or discovered, beyond the ones which we are guided to by the 
EGRET (in the future: GLAST) catalogues. Many sources will require detailed study, for which 
small-angle telescopes like MAGIC or ECO-1000 are ideally suited. 

This report is structured as follows:
The physics arguments are presented in the next section \ref{sec2}.
Sections  \ref{sec_sigbkg} and \ref{sec3}
discuss the characteristics of the gamma signal and the expected
sources of backgrounds, and the conceptual choices for a low-energy telescope, 
based on simulations of gamma ray induced showers and background. 
We concentrate on a comparison between the present
MAGIC and both a MAGIC with a high-QE camera and an ECO-1000 telescope, 
also with a high-QE camera. In section \ref{sec:analysis}, we discuss specific problems 
associated with opening the low energy sector.
We then follow, in section \ref{sec4}, with a discussion of the
practical implementation, viz. a crude work plan.

\section{\label{sec2}The Physics} 
We discuss in this section the potential in both astrophysics and fundamental physics
by observing gamma rays in the energy range from 5~GeV up, thus accessing a range
not observed with sufficient sensitivity by satellite experiments, and creating an 
overlap with these devices. In short, we try to answer the question 
{\it why go for a low energy threshold?}

The case for low threshold concerns a whole plethora of subjects, among them
\begin{itemize}
 \item  Supernova remnants and plerions: 
observations at low energy will help in
discriminating between the various acceleration mechanisms assumed to be at the 
origin of VHE gammas. 
 \item  Pulsars: The known pulsars have cutoff energies of their pulsed 
emission in the few-GeV range, hence their detection will become possible by lowering
the IACT threshold.
 \item  Unidentified EGRET sources: an enormously 
rich field of activity for 
detailed studies, possible with modest observation times on nearly half of the 
observable EGRET sources.
\item  Fazio-Stecker relation: observation at lower threshold implies access to larger
redshifts, and multiple AGNs up to redshifts 2 will help in determining the FSR (gamma
ray horizon) determined by the infrared background light, 
resulting in better understanding of the cosmological evolution of galaxy formation, 
and the role of dust-obscured galaxies.
 \item  Quantum gravity: the search for effects using time delays 
as a function of energy, in a large number of sources, will improve with lowering the
threshold, due to an increased number of of potential time-variable gamma ray sources, 
and the larger distances to them.
 \item  Dark matter: the allowed space for the many theoretical
models of dark matter
may well be restricted by a systematic search for neutralino annihilations. 
The recently discovered difficulty due to low flux in the expected sharp energy peak
from neutrino annihilations, can only be overcome by lowering the threshold
to identify the continuum signature.
 \item  Nearby galaxies: their expected steep energy spectrum 
makes observations at low energy a particularly good argument, as they allow enough flux
to be detected in the gamma ray domain.
\end{itemize}

The details concerning physics becoming possible with a low energy threshold
gamma telescope are given in the following paragraphs. 
In all cases, the argument can be
made that approaching the physical limit of detectability for gammas ($\sim$2-3~GeV)
will give access to information presently not available from any instrument, and will
in many cases allow to discriminate between competing theoretical models. 
We are confident that these arguments will be
borne out, at least in part, already by the results from MAGIC, 
which has started taking data in late 2003.

\subsection*{Fundamental Physics and Exotics}

{\bf Fazio-Stecker relation}

Gamma rays from distant sources interact in the 
intergalactic space with infrared 
photons, and are substantially attenuated. The flux reduction depends on 
the photon energy and on the distance (redshift) of the source; the 
Fazio-Stecker relation (FSR \cite{FSR}, or Gamma Ray Horizon) is 
commonly defined as that energy for each redshift for which the 
photon flux is reduced by a factor e. 

The detection of sources beyond the FSR is 
extremely  difficult, due to the strong flux suppression. This has been
a strong argument for lowering the energy threshold for the present 
generation of instruments (MAGIC). If their performance is as predicted, 
sources up to redshifts z $\sim$2 will become accessible, at least at 
small zenith angles. Observations 
at larger zenith angles  and uncertainties in the models of the extragalactic 
background light (EBL) \cite {dejag},
seem to suggest that this goal can be reached only with difficulty. 
Lowering the threshold to few GeV will provide enough lever arm to measure
with precision the exponential energy cutoff due to cosmological 
absorption, which reaches about 20-40~GeV. Also, 
measurements of cosmological parameters through
a determination of the FSR \cite {blanch} should become possible
with a large sample of AGNs distributed over a range of redshifts up to 
z $\sim$4.

 {\bf Quantum Gravity}
  
Astronomical objects have been proposed as good laboratories to study fundamental 
physics, not accessible by accelerator facilities on Earth because of the huge 
energies and masses necessary to show measurable effects. In particular, Gamma 
Astronomy and Imaging Air Cherenkov Telescopes (IACTs) have been proposed
to observe possible effects due to a quantum formulation of gravity \cite{qg1}.

Formulations of Quantum Gravity contain naturally quantum fluctuations of 
the gravitational vacuum, and lead to an energy-dependent velocity for electromagnetic 
waves. In other words, gammas of different energy produced simultaneously in an 
extragalactic object, arrive on Earth at different times due to their propagation through 
the gravitational vacuum. This Quantum Gravity time delay allows to measure 
Quantum Gravity effects. 
The speed dispersion relation might be quite different for each Quantum Gravity 
formulation. The differences are small in all possible models, and can be 
studied from a phenomenological point of view only. Typically,
\par
\begin{center}
$\Delta t = \eta (E / E_{QG})^{\alpha}$
\end{center}
\par
where $E_{QG}$ is an effective energy scale of Quantum Gravity (i.e. must be close to the 
Planck mass), $\alpha$ is the first non-zero leading order, and $\eta$ 
is proportional to the gamma path 
from the source to the observer, to a good approximation 
the distance to the source.

The equation above predicts the largest time delays for the most 
energetic gammas. A low energy threshold 
will thus not help directly in this measurement, but will have indirect effects.
An energy threshold at zenith of few GeV will allow to 
have energy thresholds of tens of GeV at medium-to-large zenith angles, leading to 
effective areas larger by an order of magnitude, and hence will 
collect 10 times more gammas than the 
present IACTs in the same energy range. One of the 
major problems in the measurement of time delay, viz. the lack of gammas at 
medium-to-large energies, can thus be overcome. Also, any time delay effect 
must be studied over a wide range 
of redshifts, in order to disentangle any Quantum Gravity effect from 
source-dependent time delay emission effects. For this reason, a large number of 
sources at different redshifts must be detected and measured 
with some precision. A low-energy threshold detector as 
ECO-1000 will be an ideal instrument for such studies.

{\bf Dark Matter Search}

Cosmology provides strong arguments, in explaining the observed Universe Structure, 
that some 25\% of the energy in the Universe is in the 
form of Dark Matter, most of it "cold" and much of it as non-barionic relic 
"Weakly Interacting Massive 
Particles" (WIMPs). WIMPs should constitute a 
clumpy halo around galaxies, and concentrate around the galaxy center and in dwarf 
spheroidal satellites (dSph), identifiable by having a large mass-to-light ratio 
\cite{darkm1}.

The Standard Model of Particle Physics has 
no candidate for WIMPs, and the nature of Dark Matter 
can only be explained by going beyond \cite{darkm2}. 
Supersymmetric Theories are among the most popular extensions to the Standard Model.
They assume a symmetry between the fermionic 
and bosonic degrees of freedom, providing a Grand Unified framework for the 
fundamental coupling constants in agreement with the high precision collider 
measurements, and giving an elegant solution to some fundamental theoretical 
loopholes of the Standard Model. Supersymmetry (SUSY) also provides a quite natural 
candidate for the WIMP, the Neutralino. SUSY particles, if they exist, 
shall be detected at the Large Hadron Collider (LHC).
The Neutralinos could be the Lightest Supersymmetric Particles (LSP) and, if R-parity (a 
quantum number associated to the new SUSY particles) is conserved, should, 
therefore, be stable. 
They are weakly interacting since they are a mixed state of the SUSY spin ½ partners of 
the neutral electroweak gauge bosons (Gauginos) and the two neutral Higgs bosons 
which are mandatory in SUSY theories (Higgsinos).

Relic Neutralinos may, in principle, be detected 
directly through their elastic scattering while they 
impinge on instruments on Earth. In addition they might be 
indirectly detected by their annihilation through different 
channels, producing finally high energy gammas and neutrinos \cite{darkm3}.
Several annihilation channels lead to the production of high energy gamma rays in the 
range of present and future Cherenkov telescopes; of these, 
$\chi\chi \longrightarrow \gamma\gamma $ and $\chi\chi \longrightarrow Z \gamma $ 
would provide ideal observational results (monochromatic annihilation lines), but are 
strongly suppressed. Instead, the most probable reaction is
$\chi\chi \longrightarrow jets \longrightarrow n \gamma $, which could be observed as
an energy distribution in $\gamma$s different from the power laws 
characteristic for the cosmic 
acceleration mechanisms.

Supersymmetric Theories have multiple free parameters, but the requirement that 
they provide Grand Unification and a satisfactory answer to the theoretical clues of the 
Standard Model, together with the constraints from the non-observation of their effects 
up to now, restricts their parameter space. A set of parameter space benchmark points (A-M) incorporating all these 
constraints have been suggested by the SUSY-search community, and provide a clear 
framework to make predictions for future SUSY hunting at accelerators and in other 
detectors \cite{darkm4}.

\begin{figure}[t]
\begin{center}
\epsfig{file=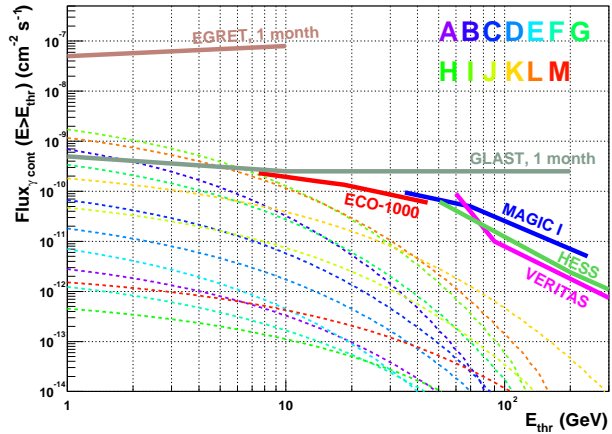,width=9cm} 
\end{center}
\caption{\small \it Integrated gamma flux as function of photon energy threshold, 
for gammas 
produced by relic annihilations in the galactic center, assuming a moderate halo 
parameter of J=500. A to M are the Post-LEP SUSY benchmark points, extracted 
from \cite{darkm5}. The point source flux sensitivities for several 
gamma ray detectors are also shown.}
\label{fig:dm1}
\end{figure}

Figure \ref{fig:dm1} shows flux predictions for these benchmark points, as a function 
of the photon energy threshold for gamma rays 
produced by relic Neutralino annihilations in the galactic center \cite{darkm5}. 
A moderate halo 
density profile J=500 has been assumed; for particularly cuspy halo 
models, such as those in \cite{darkm6}, the model fluxes are higher by two 
orders of magnitude, leading to detectable signals in GLAST andi MAGIC. 
There is no clear feature in the spectrum, the only way of 
discriminating against the astrophysical background is to study the energy 
spectrum within at least one decade in energy. 
For that, a Cherenkov telescope such as 
ECO-1000 with low threshold and high flux sensitivity is mandatory.
The fact that the galactic center is at large zenith angle for Northern Observatories,
argues again for a low energy threshold in the search for Neutralino 
annihilations at the galactic center.

{\bf Gamma Ray Bursts}

Nearly 3000 Gamma Ray Bursts (GRBs) have been observed by BATSE, but 
the phenomena causing them are still a mystery.  Even the successful 
simultaneous observation of some GRBs
at different wavelengths has not helped to find a favorite among the numerous 
existing theoretical models. The GCN (GRB Coordinates Network) is an attempt to
encourage multiwavelength observations, by distributing information
coming mainly from satellites, in real time.  We know
at least that there exists an \emph{afterglow}, which 
follows immediately the burst itself, normally lasting longer for observations at lower 
energy. GRBs last between a few seconds and minutes, the X-ray emission typically 
runs on a scale of days, and the optical one even on a scale of weeks.

Some BATSE observations (less than a percent) 
were complemented by EGRET, at $\sim 1\,\mathrm{GeV}$ of energy;
however, both the field of view (FOV) and 
the sensitivity of EGRET did set severe limits. Encouraging is the fact that
two of the EGRET-detected GRBs (GRB930131 and GRB940217) 
lasted longer inside the EGRET energy window than the observation in BATSE, so that
an afterglow at higher energies can not be excluded, for some GRBs\footnote{The
existence of exceptionally long GRBs, where the gamma ray emission lasts 
clearly longer than the one in the hard X-ray range, is particularly challenging 
for theoretical models, that have to deal with a continuous acceleration process 
at a substantially higher energy than that of the prompt emission}.

ECO-1000 will be in a better position than EGRET: the limited FOV can
be compensated by the fast repositioning system, and the sensitivity is 
greatly increased by the calorimetric observation of gamma rays typical of 
Cherenkov instruments.

\subsection*{Astrophysics}

{\bf Supernova Remnants}

Shell-type supernova
remnants originating in core-collapse supernovae,
have long been suggested to be the sites
for cosmic ray acceleration below 100~TeV, 
mainly on the basis of general energetics 
arguments \cite{blan}. We know from their synchrotron, 
radio and X-ray emission that electrons 
are accelerated to TeV energies. 

However, there is no direct evidence for proton 
acceleration. A possible signature of proton 
acceleration would be the spectrum of
$\pi^{\circ}$~decay from collisions
of cosmic ray protons and nearby matter
like high density
molecular clouds  \cite{drur}. A number of shells have 
been observed by EGRET at 0.1-10~GeV energies 
\cite{espo}, \cite{torr},
and by IACTs at TeV energies \cite{tani},
\cite{mura}, \cite{ahar}; nevertheless,
the origin of this radiation
remains uncertain, due to the contamination
of $\gamma$-rays produced by leptonic
processes (Bremsstrahlung or Inverse-Compton).

A low energy threshold Cherenkov telescope will be
instrumental in disentangling both emissions,
either through the different spectral shape
or through resolving $\gamma$-ray features that
coincide with high density matter regions.
A low threshold and increased sensitivity will allow spectral
studies in the $>$10~GeV range, where the different
mechanisms are expected to show different
spectral shapes \cite{gais};
it also may allow 
to pin down the exact positions of possible 
spectral cutoffs. Increased 
photon statistics at energies around 1~GeV 
may allow to reject pulsars.
An angular resolution close to 0.1$^{\circ}$
at energies below 10~GeV may allow to discriminate 
regions of enhanced matter density in direct
interaction with the SNR shock, point source 
emission from pulsars, extended emission
coincident with plerions, or regions of low
density where the emission is most probably due
to leptonic processes.

{\bf Plerions}

Plerions or Pulsar-Wind Nebulae are SNRs
in which a pulsar wind injects energy into
its surroundings. A bubble is inflated 
out to a radius where it is confined by the
expanding shell, as already suggested by 
\cite{rees} for the Crab Nebula. 
Particle acceleration is expected in the
wind termination shock.

\begin{figure}[t]
\centering
\epsfig{file=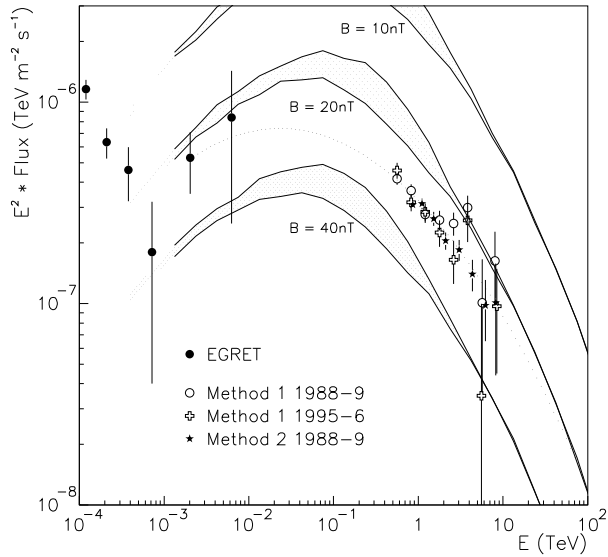,width=8cm}
\caption{\small \it Predicted IC spectrum
according to the references in the text, 
compared to the EGRET, Whipple (method 1 and 2) 
and CANGAROO observed spectra}
\label{fig:jc2}
\end{figure}                                               

Figure \ref{fig:jc2} shows the Crab plerion spectrum 
measured by EGRET at energies up to 10~GeV, 
and by Whipple and CANGAROO at TeV energies,
along with the predicted spectra for different
models \cite{hill}, \cite{deja}, \cite{atoy}. 
The VHE emission 
is probably due to Inverse-Compton of $\Gamma >$ 10$^8$
electrons. The observed synchroton X-ray emission 
confirms the existence of these extremely high 
energy electrons, while the dynamics of the
particle flow only yield $\Gamma \sim$ 10$^6$
for the postshock region. 

A precise measurement
of the spectrum in the 1-100~GeV energy range
is crucial to constrain the model parameters
and to ascertain if another source of photons
is necessary, possibly Bremsstrahlung 
from dense regions of gas.

{\bf Pulsars}

The observation of gamma ray pulsars in the GeV domain is of special 
interest: in 
this range, the EGRET pulsar spectra have a cut-off, 
and observing the differential spectrum of a gamma ray pulsar 
will discriminate between the polar cap and outer gap models for
emission. The two models 
predict different cut-off energies, below 50~GeV for the polar cap,
up to 100~GeV for the outer gap model \cite{vict1}, \cite{vict2}.

GLAST will extend the exploration of the gamma ray sky up to 300~GeV, 
and will have, at GeV energies, a sensitivity many times higher than 
EGRET, but its limited detection area will 
restrict its capability in the high-energy range. MAGIC will observe
pulsed emissions in the non-imaging mode, and not far from its energy threshold.
The expertise and know-how of MAGIC will be a key factor for exploring
ECO-1000 with its absolute energy limit 
of 2 to 3~GeV, close to the threshold energy for 
secondary electrons to radiate Cherenkov light in the 
upper atmosphere. In ECO-1000, pulsars can be measured in both 
imaging and non-imaging 
mode. The first estimations of observation times and our calculation of
collection areas show how much ECO-1000
can be superior to MAGIC:

\begin{tabular}  {c c c c c c c}
Object (pulsar) & K $\times 10^8$ & $\Gamma$  &  
$T_{ECO-1000}$  &   $T_{MAGIC}$\\
   &  $cm^{-2}s^{-1}GeV^{-1}$  &  & [minutes]  &  [hours]  \\
Crab            &  24  &  2.08  &  9  &  1.2  \\
Geminga         &  73  &  1.42  & 26 & 24.7  \\
PSR B1951+32   &  3.8 &  1.74  &  44 &  3.2  \\
PSR J0218+4232 &  1.9 &  2.01  & 1860 & 39.5 \\
PSR J1837-0606 &  5.5 &  1.82  &  90  & 17.1 \\
PSR J1856+0113 &  7.4 &  1.93  &  84  & 17   \\
PSR J2021+3651 & 11.5 &  1.86  &  25  & 49   \\
PSR J2229+611  &  4.8 &  2.24  &  852 & 200  \\
\\
\end{tabular}
\par
In this table, we estimate the observation times to 
achieve a $5\sigma$ significance.
K is the monochromatic 
flux at 1~GeV, and $\Gamma$ is the spectral index of the source. 
The improvement in detection time is substantial (collection areas for MAGIC and 
ECO-1000 are shown in figure \ref{figcollar}).

{\bf Active Galactic Nuclei}

Galaxy formation and evolution in the early universe is one of the main open questions
of extragalactic astronomy. The conventional understanding is based on a co-evolution
of galaxies and super-massive black holes, which power the AGN phenomenon. Recent 
observations, however, seem to challenge this understanding, by showing early star and 
galaxy formation, compared to a later evolution of AGNs. 

Gamma ray observations of AGNs at all redshift will help in understanding the evolution 
of AGNs and their central engines, super-massive black holes. 
The high-energy gamma radiation emitted at close distances from the black hole is also a 
fundamental ingredient for multi-wavelength studies. The low threshold of ECO-1000 
will enable us to study the gamma ray emission of a large population of AGNs, up to 
redshifts above z~=~2, as the lower-energy gammas can reach us largely unabsorbed by 
the meta-galactic radiation field.

\begin{figure}[t]
\begin{center}
\epsfig{file=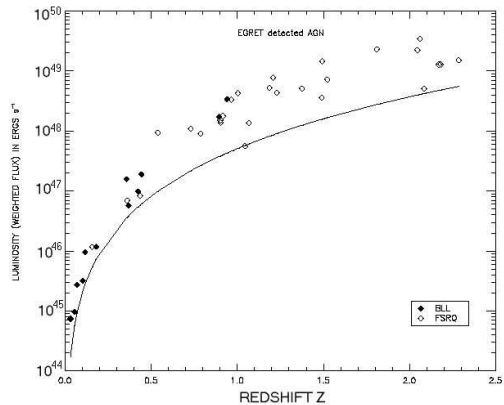,width=7cm}
\caption{\small \it AGN luminosity at different redshifts, as measured by EGRET}
\label{fig:agn_egret}
\end{center}
\end{figure}   

{\bf Microquasars and X-ray Binaries}

X-ray binaries provide a nearly ideal 
laboratory to study nearby objects of 
strong gravity, presumably black holes of a few solar masses, 
remaining from the core-collapse of 
short-lived massive stars through their 
high-energy emission.  The
sources are natural candidates for gamma rays up to at 
least GeV energies. The magnetic field frozen 
into the ionized, differentially rotating accretion disk 
around the black hole in a binary system, can twist 
and reconnect to release the energy stored in the 
magnetic field into the kinetic energy of coronal 
particles.  Short-lived, large-scale electric fields can 
accelerate charged particles up to high energies, 
scaling from numerical simulations of magnetic 
reconnection in the solar magnetosphere, the Earth's  
bow shock, and the geomagnetic tail \cite{hosin}. 
Jets forming in radio-emitting X-ray binaries 
and the gamma ray emission observed 
with EGRET up to 10~GeV are indicative of particle 
acceleration processes occurring at shocks in 
these super-Alfvenic flows.  

Among the X-ray binaries, 
microquasars form a particularly interesting 
subclass, exhibiting relativistic jets as inferred 
from superluminal motion of radio knots. Nonthermal 
radio-to-X-ray emission extends through the 
inverse-Compton process into the GeV-TeV domain 
and should be observable \cite{georg}.   In particular, microblazars 
(microquasars with their jet axes aligned roughly 
to the line of sight of the observer) promise to be an 
interesting target, for studies of short-term variability.  
Microblazars migth even be observable from 
nearby galaxies due to their Doppler-boosted flux. 
 
One important question concerns the fraction of 
cosmic rays accelerated in blazars, taking into account 
the beamed-away fraction of the sources.  If 
microquasars can be established as powerful sources 
of gamma rays in the $>$10~GeV range, a new 
paradigm for cosmic ray 
acceleration in explosive sources such as 
Supernovae (and their remnants) and Gamma Ray Bursts 
might emerge, in which quasi-steady 
sources such as Radio Galaxies, BL LACs, and 
Microquasars are responsible for most of the 
cosmic rays \cite{heinz}. 

{\bf Diffuse Photon Background}

The diffuse photon background may be classified according to its origin : 
The Extragalactic Background Radiation (EBR) and the Diffuse Galactic Emission 
(DGE). EBR is essentially isotropic, and is well established 
up to energies of $\sim$50~GeV \cite{sreekumar98}. In the energy range
from 30~GeV to 100~TeV, a large part of the EBR may be due to the direct 
emission from Active Galactic Nuclei (AGN), which have not yet been resolved
\cite{stecker96,stecker01}. 
Direct measurements of the DGE exist up to energies of $\sim$70~GeV 
\cite{hunter97}. The data 
are explained by the interaction of cosmic-ray electrons and hadrons with
the interstellar radiation fields and with the interstellar matter
\cite{bertsch93,strong00}. The production mechanisms are synchrotron radiation 
of electrons, high-energy 
electron bremsstrahlung, inverse Compton scattering with low-energy photons,
and $\pi^0$ production by nucleon-nucleon interactions. 

New measurements of the gamma radiation (either diffuse or from point-like 
or extended sources)
in the 10 to 300~GeV energy region will help to better understand the 
different sources of the diffuse gamma ray background :

\begin{itemize}
\item By the detection or identification of new AGNs one will test
the unresolved-blazar model for the origin of the apparent EBR. 
The basic assumption is an average linear 
relationship between gamma ray and radio fluxes. Such a relation is suggested
if the same high-energy electrons are invoked as the source of both the radio 
and gamma ray emission \cite{jorstad01}.
 \item A deep observation of new blazars, with measurement 
of redshift, energy spectrum, and cutoff energy, will allow a more
reliable determination of the collective luminosity of all gamma ray blazars.
A good knowledge of this contribution is a precondition for future tests
of the predictions of the cascading models for the EBR.
 \item The detection or identification of new SNRs and pulsars will allow 
better estimates of their contribution to the DGE.
 \item Measuring the absolute diffuse gamma ray flux (mainly DGE)
as a function 
of energy, galactic latitude, and longitude, will yield new 
information about the origin and propagation of galactic cosmic rays, and about
the spatial distribution of the interstellar matter, radiation and magnetic 
fields.
\end{itemize}

{\bf Unidentified EGRET Sources}

The EGRET experiment (1991-2000) has given us the first detailed view of the 
entire high-energy gamma ray sky. Many of them, including blazars,
have not been detected by IACTs.
Of the 11 VHE gamma ray
sources known today, 6 are Blazars and 5 SNRs/Plerions.
About half of these have been observed by EGRET. 

We can thus classify the VHE sources: 
(a)~sources with a steep cut-off 
which are detected by EGRET, but become unobservable above a few 100~GeV, 
(b)~flat spectrum sources like Mkn501,
which only become observable at energies well above 100~MeV, and 
(c)~intermediate cases like Mkn421 or the Crab Nebula.
Most known sources must belong to type (a), given the way observation 
technology has developed.

The number of detectable sources decreases rapidly with rising energy threshold,
even though the point source sensitivity already of present Cherenkov telescopes 
is many orders of magnitude better than that of EGRET.
It seems that
the universe becomes abruptly much darker above a few 100~GeV. 
In other words, the cut-off of most high-energy source 
spectra seems to take place in the
range 1 - 200~GeV. The first decade of energy in this range has 
been covered by EGRET and
57 sources have been detected \cite{onegev}. 
The range from 10 to 200~GeV, however, has never been explored
until today, and it is this ``gap'' that forms the major 
incentive behind more sensitive Cherenkov telescopes like ECO-1000.

We have estimated observation times ($5\sigma$) for the observable EGRET
sources, and found that $\sim$40\% of them can be studied by ECO-1000 in 
very few hours or less. The accessible sources touch all fields of 
high-energy astrophysics, and their observation will much contribute to 
constraining existing models in the energy domain where cut-offs set in.

{\bf Galaxy Clusters}

Galaxy clusters contain a hot, intracluster medium (ICM),
which acts as a storage volume for cosmic 
rays escaping from galaxies in the cluster, 
and from active galactic nuclei \cite{gabic}. The 
total energy  and the spectrum of the stored 
relativistic particles is unknown, and  gamma ray 
observations would provide important clues 
about their origin.  It is important to distinguish this 
emission component from others, possibly related 
with the annihilation of supersymmetric dark 
matter particles. Due to the expected steepness 
of the cosmic ray spectrum in the ICM, it is 
important to achieve a low gamma ray threshold. 
Other suspected sources of relativistic particles are 
the supersonic motion of galaxies through the ICM 
and the accretion of material from metagalactic 
space onto the cluster which induce the formation 
of gigantic shock waves possibly accelerating 
particles up to the highest observed energies \cite{kehse}. 
The observed nonthermal radio-to-UV emission in clusters 
ensures the production of gamma rays through the inverse-Compton 
scattering process.  There is also a contribution of 
gamma rays due to a calorimetric effect based on 
the pair production process.  The gamma rays 
from sources residing in the cluster and above 
threshold for pair production with microwave 
background photons convert into pairs, which 
subsequently scatter microwave background photons 
to higher energies shaping a gamma ray halo 
\cite{aha2}. 

{\bf Starburst Galaxies}

\begin{figure}[t]
\centering
\epsfig{file=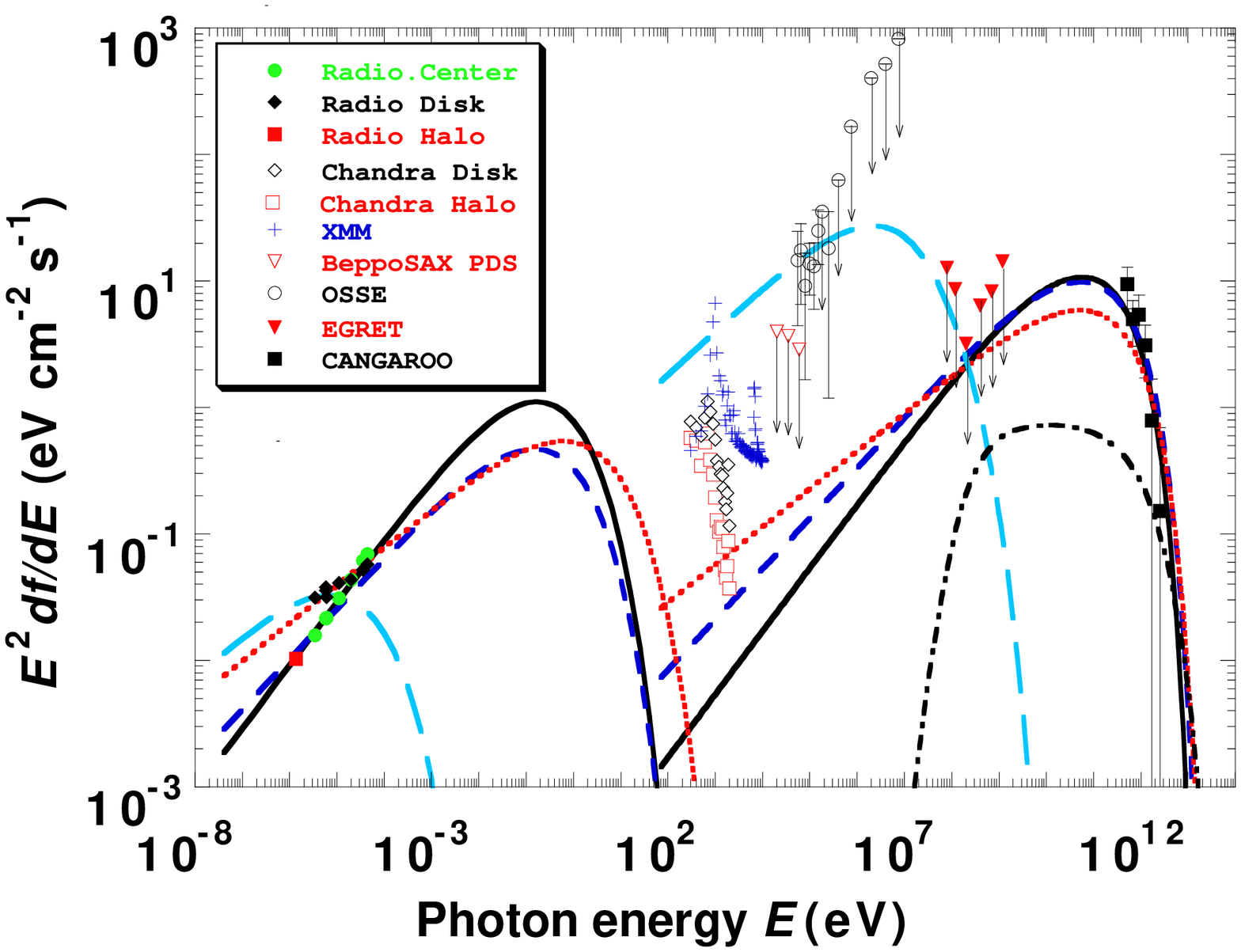,width=9cm}
\caption{\small \it Multiwavelength spectrum of the
starburst NGC~253, along with a model of the
electron IC emission in the halo (for two different
electron spectra, solid black and dashed blue),
in the disk (red), IC localized in the galactic center 
(dashed cyan), and $\pi^{\circ}$~decay}
\label{fig:jc3}
\end{figure}

The Cherenkov telescope CANGAROO has recently reported
the first detection of a normal spiral galaxy
at TeV energies \cite{itoh}. NGC~253 is a nearby
($\sim $2.5~Mpc) starburst galaxy, in which a
high cosmic ray density and non-thermal emission are
expected. The source is extended
with a width of 0.3-0.6$^{\circ}$ (corresponding
to 13-26 kpc), and temporally steady over two years.
It can be considered the first of a new class of extragalactic
objects, clearly different from the other
observed extragalactic TeV emitters (AGNs of the
BL Lac class). The TeV $\gamma$-rays may come from
hadronic or leptonic processes originating from the
cosmic ray density in a starburst (assumed high). 
Figure \ref{fig:jc3} shows the multiwavelength
spectrum of NGC~253 and estimates of the hadronic and
IC emission produced by disk and halo
electrons (from \cite{ito1}). 

 Romero et al. \cite{rome} have advanced an alternative
explanation based on hadronic processes in the
core of the galaxy. They suggest that proton illumination
of the inner winds of massive stars could produce
TeV $\gamma$-rays without the unobserved
MeV-GeV counterpart. The enhancement in cosmic ray 
density would be produced by collective effects
of stellar winds and supernovae.

Precise spectral
measurements by ECO-1000 in the range 1-100~GeV will allow to
constrain the different models; the angular resolution
around 0.1$^{\circ}$ will help to localize
the source of emission at low energies.
Nearby starburst galaxies are ideal targets
for this kind of studies.

{\bf Nearby Galaxies}

Nearby galaxies such as M31, M82, Arp~220, and Cen~A 
are representative of normal spiral 
galaxies, starburst and merger galaxies, 
and active galactic nuclei.  Owing to their vicinity, the 
morphology of these galaxies and their 
multifrequency properties have been studied in great 
detail, but lack information in the 5~GeV-300~GeV region.  
In normal galaxies, star formation 
and hence the production of collapsing massive 
stars with associated gamma ray production in 
GRBs, supernovae and their remnants can be 
probed with a low-threshold IACT  like ECO-1000 \cite{itoh,rome}.  
The low threshold of 5~GeV is important to observe enough flux for their 
detection, since the spectrum of cosmic rays (and hence gamma rays) 
is expected to be very steep
(-2.75 in our Galaxy).  

{\bf The Galactic Center}

The Galactic Center (GC) region, excepting the famous Sgr A*, contains many unusual 
objects which may be responsible for the high energy processes generating gamma rays. 
The GC is rich in massive stellar clusters with up to 100 OB stars
\cite{morris}, immersed in a dense gas within the volume of 300 pc and
the mass of $2.7\times 10^7 M_\odot$, young supernova remnants e.g. G0.570-0.018 or
Sgr A East, and nonthermal radio arcs.

In fact, EGRET has detected a strong source in the 
direction of the GC, 3EG J1746-2852 \cite{mayer}, which has a 
broken power law spectrum extending up to at least 10~GeV, with the index 1.3 below the 
break at a few GeV. If in the GC, the gamma ray luminosity of this source is very large
$\sim 2\times 10^{37}$ erg s$^{-1}$, which is equivalent to $\sim$ 10 Crab pulsars.
Up to now, the GC has been observed at TeV energies only by the HEGRA Collaboration
\cite{pohlh}. The upper limit has been put on 1/4 Crab (results from the 
CANGAROO-II Collaboration are expected at ICRC 2003). 
High energy gamma rays can be produced in the GC in the non-thermal radio filaments by 
high energy leptons which scatter background infrared photons from the nearby ionized 
clouds \cite{pohl}, or by hadrons colliding with dense matter.
These high energy hadrons can be accelerated by the massive black hole, associated with
the SGr A* \cite{levins}, supernovae \cite{fattuz}, or an energetic pulsar \cite{bedna2002}. 
In order to shed new light on the high energy phenomena in the GC region, and constrain 
the models above, new observations with sensitivity down to 10~GeV are necessary.

\vspace{2mm}  \par
{\bf ECO-1000 and GLAST}

The successor of EGRET, the Gamma ray Large Area Space Telescope 
(GLAST) \cite{gehrels},
is scheduled for launch in September 2006. The mission duration will
be at least 5 years, more  likely  10 years.
GLAST will have an effective collection area just under 1~m$^2$
and will be able to detect gamma rays with good energy and direction 
resolution between 0.1 and 300~GeV.
In the overlap region, from 5 to 300~GeV, GLAST and ECO-1000
will perfectly complement each other.

GLAST is a survey instrument. It has a very low background counting rate
at energies above a few GeV, and it has a large field of view.
Over its lifetime, GLAST is predicted to discover thousands of
new sources. However, these detections will be severely
photon-limited above a few GeV: From the Crab Nebula, which
is a strong source, GLAST will
detect $\approx$3500 photons above 1~GeV per year, 400 of these
photons will be above 10~GeV, and only about 23 photons will be
above 100~GeV \cite{petrydigel}. For a source significantly weaker than the Crab Nebula
it will not be possible to measure the spectrum
above 10~GeV with an accuracy 
adequate for constraining source models, even though the source
may be clearly detected. 
The low photon detection rates will also lead to a 
strong limitation for studies of short-term variability
above a few GeV.

ECO-1000, on the other hand, will have a high background rate
and relatively small field of view,
but an effective collection area four to five orders of magnitude larger
than that of GLAST. Hence, given the positions of sources discovered
by GLAST, ECO-1000 can deliver spectra above 5~GeV which have
higher accuracy and better time resolution for variability
studies. In other words, GLAST will discover the sources and 
measure their spectra and variability from 0.1 - 5~GeV while 
ECO-1000 will complete the spectra and variability studies
from 5~GeV to where the sources cut off.
GLAST and ECO-1000 form an ideal pair of complementary instruments.

The large overlap region from 5 to 300~GeV will allow
a good cross-calibration such that high-accuracy spectra
can be constructed, spanning an energy range of more than three orders 
of magnitude from 0.1~GeV to 1000~GeV or wherever the sources
cut off.

\vspace{2mm}  \par
{\bf Multiwavelength Studies of Active Galactic Nuclei (AGNs)}

The study of blazars - active galaxies radiating across the whole 
electromagnetic spectrum from radio to gamma/TeV rays - requires multifrequency, 
multiapproach studies. We should recall that a single observation in the
1990s had a large impact on the study of AGNs: the Compton 
Gamma Ray satellite Observatory (CGRO) found that a major fraction of energy 
is radiated in gamma rays. This allowed the conclusion that all 
electromagnetic frequencies from radio to gamma rays are very closely 
connected. Later, some blazars were found by ground-based IACTs 
to be extremely energetic even at TeV-energies. We can conclude that measurements
at the wavelengths becoming available to us by decreasing the energy threshold,
will contribute important knowledge to this sector.
\par
While this general framework is known, the complicated correlations between 
various phenomena at different energies make it extremely difficult to study the 
details without access to simultaneous data across the entire electromagnetic 
range. Broadly speaking, the spectral energy distribution of all radio-loud AGNs 
consists of two maxima, one from radio to UV/X-rays, produced by synchrotron 
radiation, the other from X-rays to TeV caused by inverse Compton (IC) 
radiation. No details can be understood without studying the entire spectrum.
It seems assured, then, that good
observations in the energy range, which we try to open together with GLAST, 
viz. from 5 to 50~GeV, will 
be a key contribution in understanding blazars.
\par
A crucial question to study is the nature of the seed photons. 
The accretion disk of AGNs is the most 
obvious source of photons, mainly in the UV. These photons can also be 
reflected/reprocessed by the broad line region clouds, producing an intense 
optical photon field. Farther away, dust heated to 500-1000~K is a source of 
infrared photons. All these are called external Compton (EC) scenarios, since 
the seed photons come from outside the jet. The synchrotron photons in the jet 
can also scatter from the electrons which produced them, in which case we have a 
synchrotron self-Compton scenario (SSC).
In the EC models, the high frequency emission must originate within a small 
fraction of a parsec from the AGN core, since the photon density drops rapidly 
with distance. As the synchrotron-emitting shocks typically reach their maximal 
development much further downstream, the gamma ray variations should precede the 
onset of the radio flare. Since the external photon field is independent of the 
electron density in the jet, the EC flux should change linearly with the 
synchrotron flux. In the SSC scenario the change should be quadratic, since the 
synchrotron photon density is also changing, not just the electron density. The 
SSC gamma rays can be emitted simultaneously with, or even after, 
the onset of the radio flare. 
\par
MAGIC (and future gamma-ray telescopes in La Palma) have
straight access to synchronous dedicated optical observations, being
associated with the
KVA optical telescope, situated next to the MAGIC site on La Palma. 
This is the only Cherenkov Observatory with such a facility.
Besides 
standard UBVRI-photometry, KVA will also be the only telescope in Europe with 
simultaneous polarimetric monitoring capability, a crucial asset in, e.g., 
separating thermal and non-thermal contributions to the total flux. The KVA
telescope will soon be fully automatic. 

Simultaneous 
observations in GeV/TeV and optical energies are extremely important when we 
expect to increase the number of detected sources dramatically as we reach
lower energy threshold (thus avoiding the IR-background absorption), using 
MAGIC and, eventually, the ECO-1000 telescope.
In TeV-blazars, the optical synchrotron flux (highly polarized) should be 
connected to the GeV/TeV fluxes. While contributions to the lower energy IC flux 
may come from a wide electron energy range, and from several processes, 
including thermal X-rays close to the accretion disk, only the very highest 
energy electrons can boost the seed photons to the extreme TeV energies. The TeV 
variations are, therefore, a very pure IC signal, and it should in principle be 
easy to identify the 'parent' synchrotron component on the basis of correlations 
and time lags. Observations of TeV flares have provided intriguing hints, but no 
definite answers due to insufficient simultaneous optical and GeV/TeV data. With 
joint observations of KVA and MAGIC/ECO-1000, the answers are within our reach.

\section{\label{sec_sigbkg}Signal and Backgrounds in the few GeV Domain}

The following paragraphs are the result of extensive simulations at low energies, 
up to 200~GeV, for gammas end electrons, and up to 3~TeV for protons. Throughout, 
we have used the simulation program CORSIKA (version 6.019 \cite{cor02}), 
with experiment-specific extensions; the extensions define 
three different telescopes: the existing
MAGIC telescope, a MAGIC-HQE in which all parameters remain unchanged except the
camera, equipped with high quantum efficiency photosensors, and ECO-1000, which is 
taken to have a high-QE camera with the same angular coverage as MAGIC,
and a useful mirror surface of 1000 sq.m. 
\vspace{2mm}  \par
{\bf Characteristics of $\gamma$ Showers at few GeV}

In Cherenkov experiments, the detection quality of
$\gamma$-initiated showers is directly related to the density of the 
Cherenkov photons reaching the telescope. Differences in their
characteristics are critical in discrimination against hadrons, the
dominating background.  
We have done extensive simulations, for different primary energies and 
zenith angles. The altitude (2200m a.s.l.) 
and the geomagnetic field of La Palma were used. 
We summarize in the following some characteristics of $\gamma$ showers.

\begin{figure}[t]
\begin{center}
\begin {tabular}{c c}
\epsfig{file=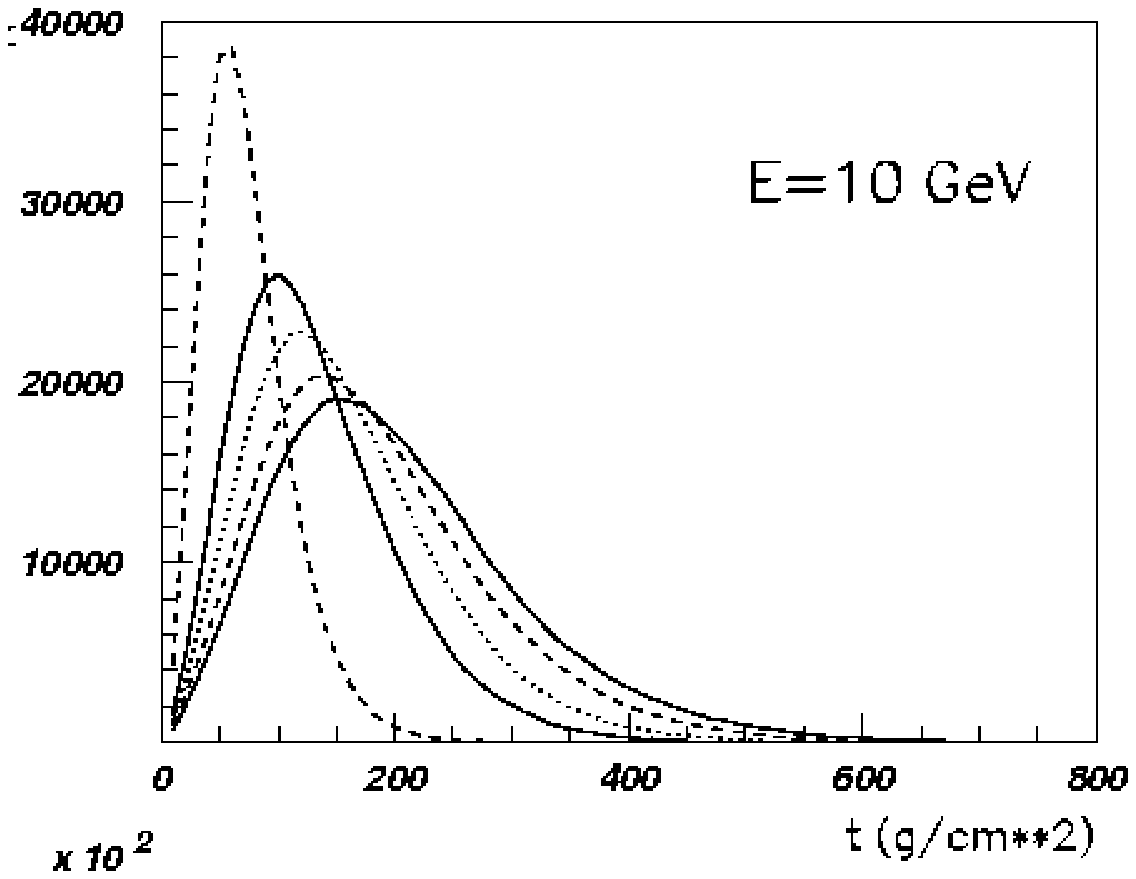,width=5cm} &
\epsfig{file=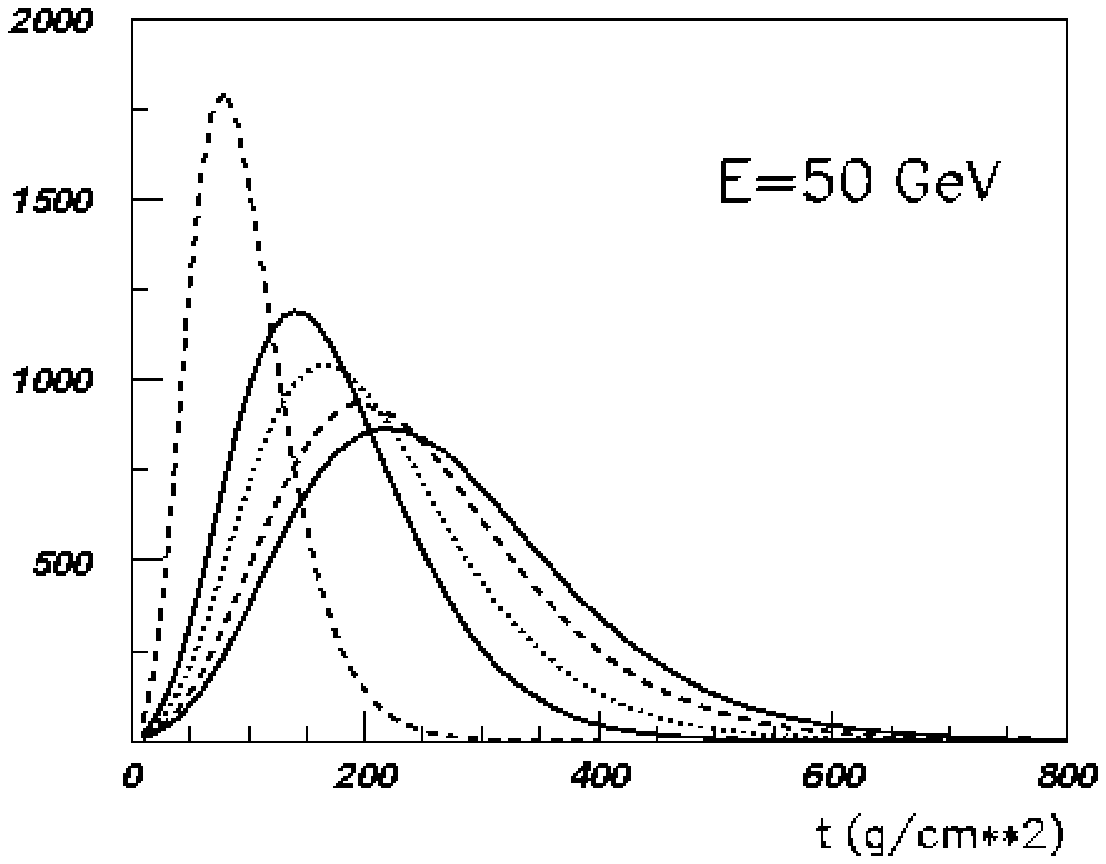,width=5cm} 
\end{tabular}
\end{center}
\caption{\small \it Longitudinal distributions
(the average number of generated Cherenkov photons) 
for different simulated primary energies and zenith angles
in $\gamma$ cascades,  along the vertical direction; 
the broadest distributions are for 
zenith angle 0, the others are for 25$^\circ$, 40$^\circ$, 
50$^\circ$, and 70$^\circ$.}
\label{fig:ds3}
\end{figure}

The average total number of  Cherenkov photons created in a $\gamma$ shower is given
in table \ref{tab:ds1}, and their longitudinal profile shown in figure \ref{fig:ds3}. 
The average total number of Cherenkov photons 
in the shower is weakly dependent on zenith angle, and it increases with 
the primary energy.

\begin{table} [t]
\begin{center}
\begin{tabular}{c c c c c c c}
E(GeV)&$zen=0^o$&$zen=25^o$&$zen=40^o$&$zen=50^o$&$zen=70^o$ \\
\\
5&0.22&0.21&0.20&0.19&0.15  \\
10&0.46&0.44&0.42&0.40&0.32 \\
20&0.94&0.92&0.87&0.83&0.68 \\
50&2.44&2.40&2.27&2.16&1.78 \\
100&5.02&4.90&4.67&4.45&3.66 \\
\end{tabular}
\end{center}
\caption{\small \it The average total number of Cherenkov 
photons created in $\gamma$ cascades, for different zenith angles,
given in millions.}
\label{tab:ds1}
\end{table}

The number of charged particles in $\gamma$ cascades, shown as a function
of depth, has a maximum depending on 
energy; it occurs at different (vertical) depths in function of zenith 
angle (ZA). The maximum value does only weakly depend on  the ZA, but the 
penetration depth at higher ZA shrinks substantially.
In low energy showers, below  100~GeV, no charged particles penetrate
to the observation level.

\begin{figure}[t]
\begin{center}
\epsfig{file=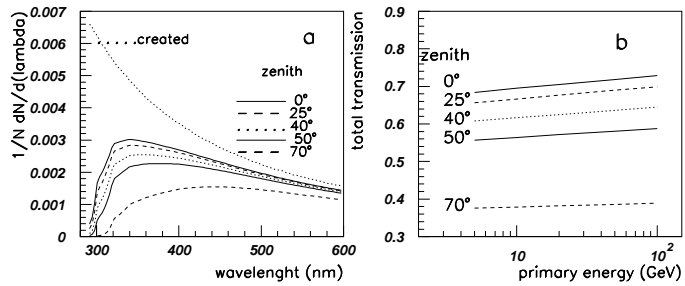,width=10cm} 
\end{center}
\caption{\small \it a) Average spectrum of Cherenkov photons after atmospheric 
absorption from 5~GeV $\gamma$-s, at different zenith angles; b)
Average total transmission (probability to reach ground) 
versus the primary energy, at different zenith 
angles}
\label{fig:ds4}
\end{figure}

All charged particles with energy above the Cherenkov threshold\footnote{the 
threshold is given by $v\ge c/n$, with n = index of refraction}
contribute to the
Cherenkov light pool. The average number of Cherenkov photons
produced in the atmosphere  is shown as a function of thickness in 
figure \ref{fig:ds3}. The distance between shower maximum and the 
observer increases with the zenith angle.

The average 
{\it spectrum} of Cherenkov light at observation level is shown 
in figure \ref{fig:ds4}a,
in comparison with the originally generated spectrum. 
As an example, 5~GeV $\gamma$ rays were choosen. The average total transmission 
(viz. the fraction of light reaching ground)
for all simulated energies and zenith angles was calculated and plotted 
in \ref{fig:ds4}b). The average 
total transmission of the light from the shower is only weakly energy 
dependent, but it decreases very clearly with the zenith 
angle.

\begin{figure}[t]
\begin{center}
\begin{tabular}{c c}
\includegraphics*[height=4cm]{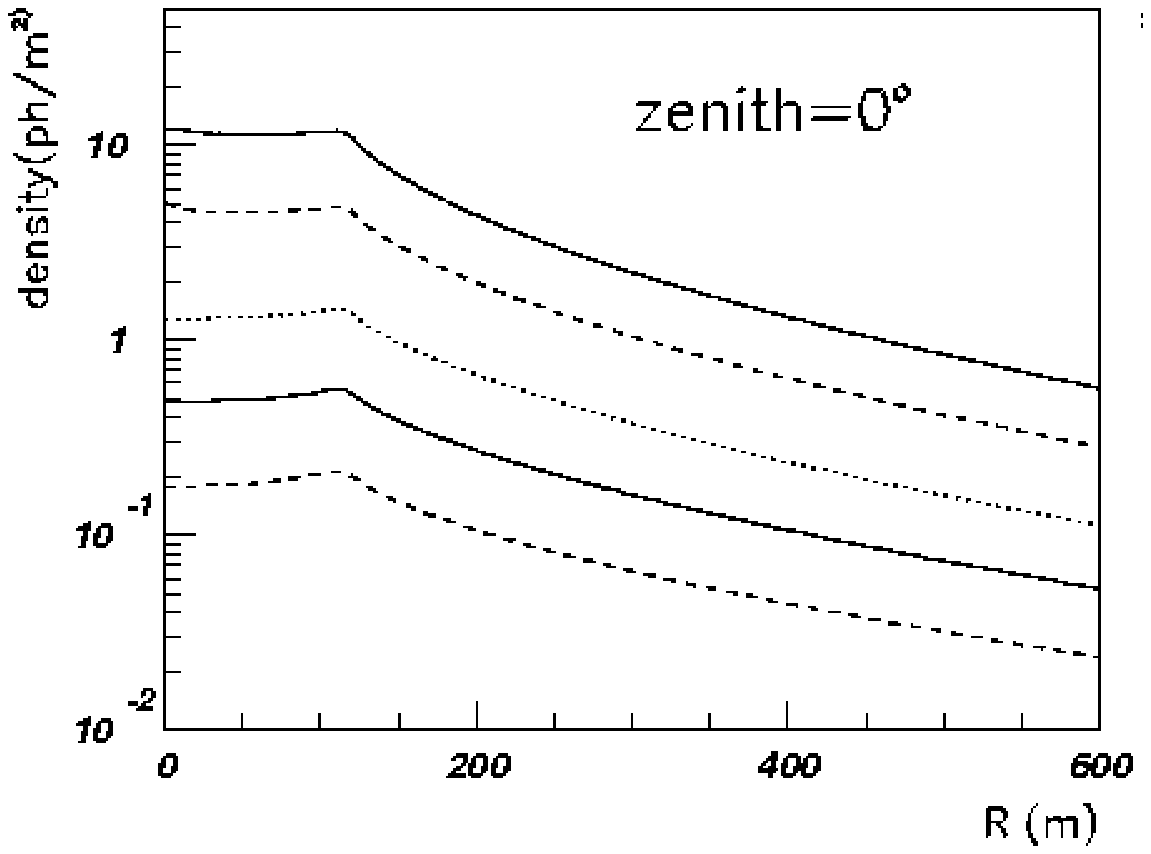}  & 
\includegraphics*[height=4cm]{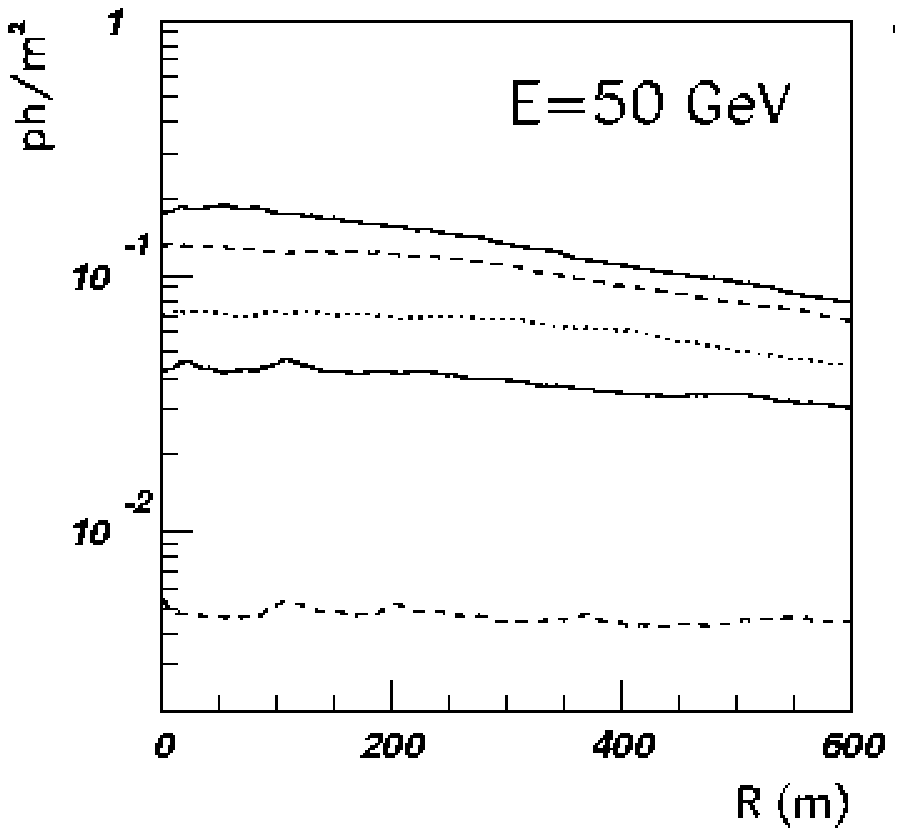}
\end {tabular}
\end{center}
\caption{\small \it Average lateral photon density; left: 
for gammas, at different energies (highest: E = 100~GeV, lowest E = 5~GeV); 
right: for protons at 50~GeV, for
different zenith angles (highest for $\Theta$ = 0, lowest for $\Theta = 70^\circ$. 
The behaviour for different E and $\Theta$ is similar for gammas and protons.}
\label{figlatdist}
\end{figure}

The average lateral density distributions were calculated in the plane 
perpendicular to the shower axis, for distances from the shower core up to 600 m.
The curves approximately scale with energy, the well-known peak position shifts
with increasing zenith angle $\Theta$
from 120~m for $\Theta$ = 0 to more than 300~m at $\Theta$ = 70$^\circ$. 
For fixed energy, we see the 
plateau of photon density decreasing with $\Theta$. Figure \ref{figlatdist}
shows the typical lateral distribution.

\vspace{2mm}  \par
{\bf Characteristics of low-energy Proton Showers}

\begin{figure}[t]
\begin{center}
\epsfig{file=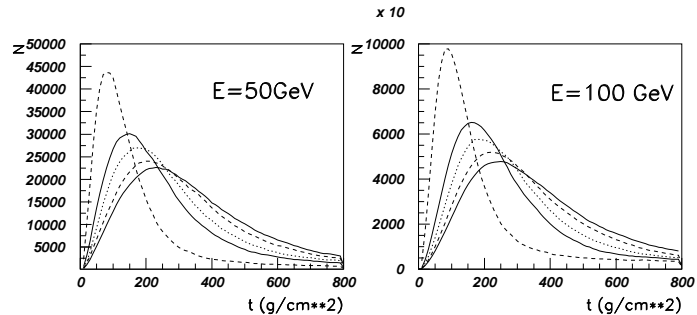,width=10cm} 
\end{center}
\caption {\small \it Longitudinal distributions (the average number of Cherenkov photons 
versus thickness) in proton showers, for two simulated primary energies and 
several zenith angles 
(the broadest curve is for zenith angle 0, the others are for 25$^\circ$, 40$^\circ$, 
50$^\circ$, and 70$^\circ$);  the thickness is measured  along the vertical direction}
\label{fig:ds8}
\end{figure}

Cosmic rays, mostly protons and Helium ions, are the dominating background
in $\gamma$ observation. Their total rate is known from measurements, e.g. 
\cite{alcaraz}.  We have extensively simulated and analyzed
proton showers.
The longitudinal development of proton showers shows that, on average, several  
charged particles reach the observation level (La Palma altitude), 
unless the energy is low (for protons, this means $<$100~GeV)
and the zenith angle large. This fact is reflected in figure 
\ref{fig:ds8}, which shows the number of created photons as a function of 
the penetration depth. This being different from gamma showers, 
the effect contributes in discriminating gammas
and hadrons, as does the narrower concentration of gamma showers.

The lateral density distribution of Cherenkov photons 
does not show a clear limitation of the light pool, with a peak, 
as do gamma showers; it simply falls with
the distance from the core axis, as shown in figure 
\ref{figlatdist}.  

\vspace{2mm}  \par
{\bf Diffuse Electron Background and the Geomagnetic Field}

\begin{figure}[t]
\begin{center}
\begin{tabular}{c c}
\includegraphics*[width=6cm]{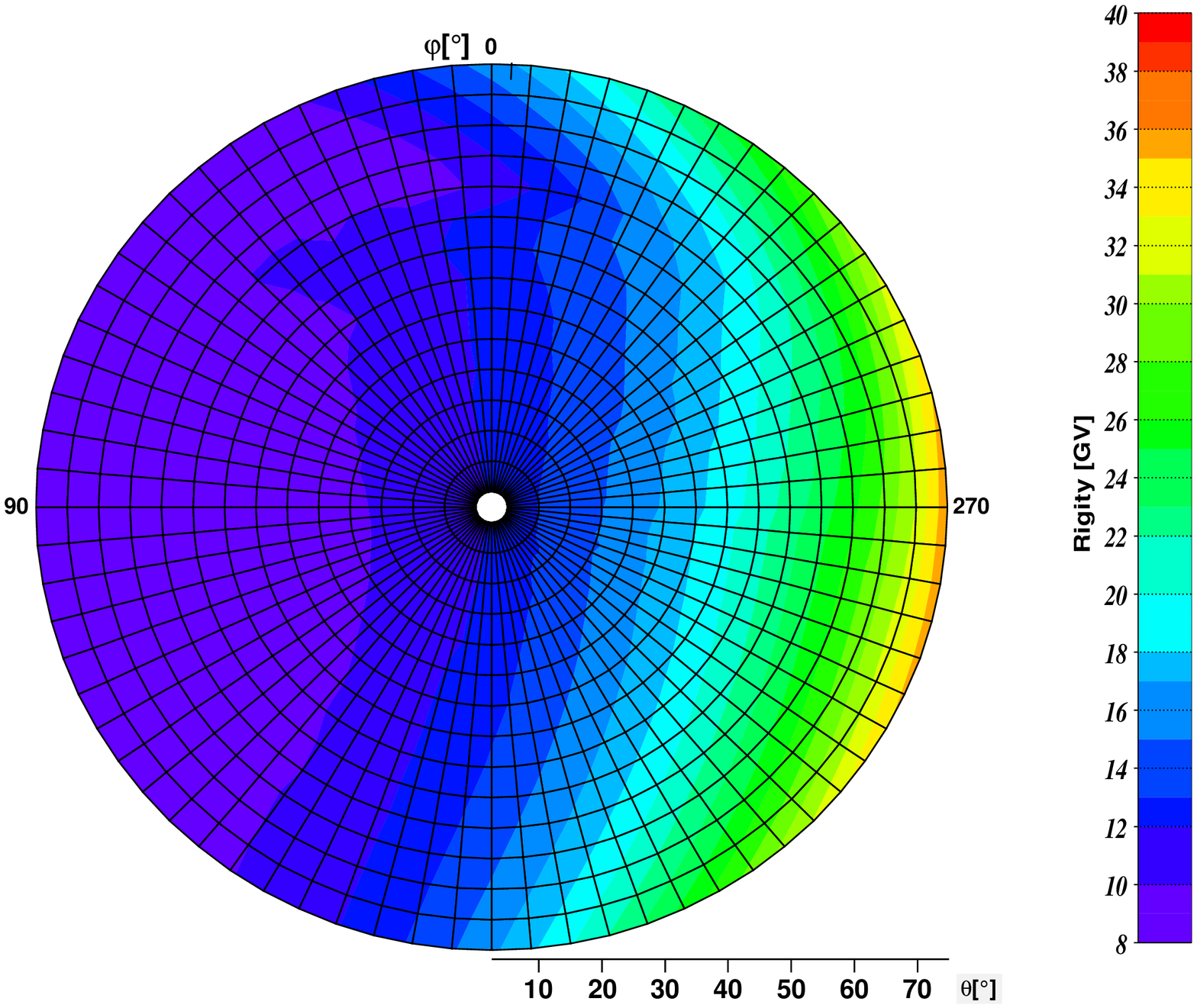}  & 
\includegraphics*[width=6cm]{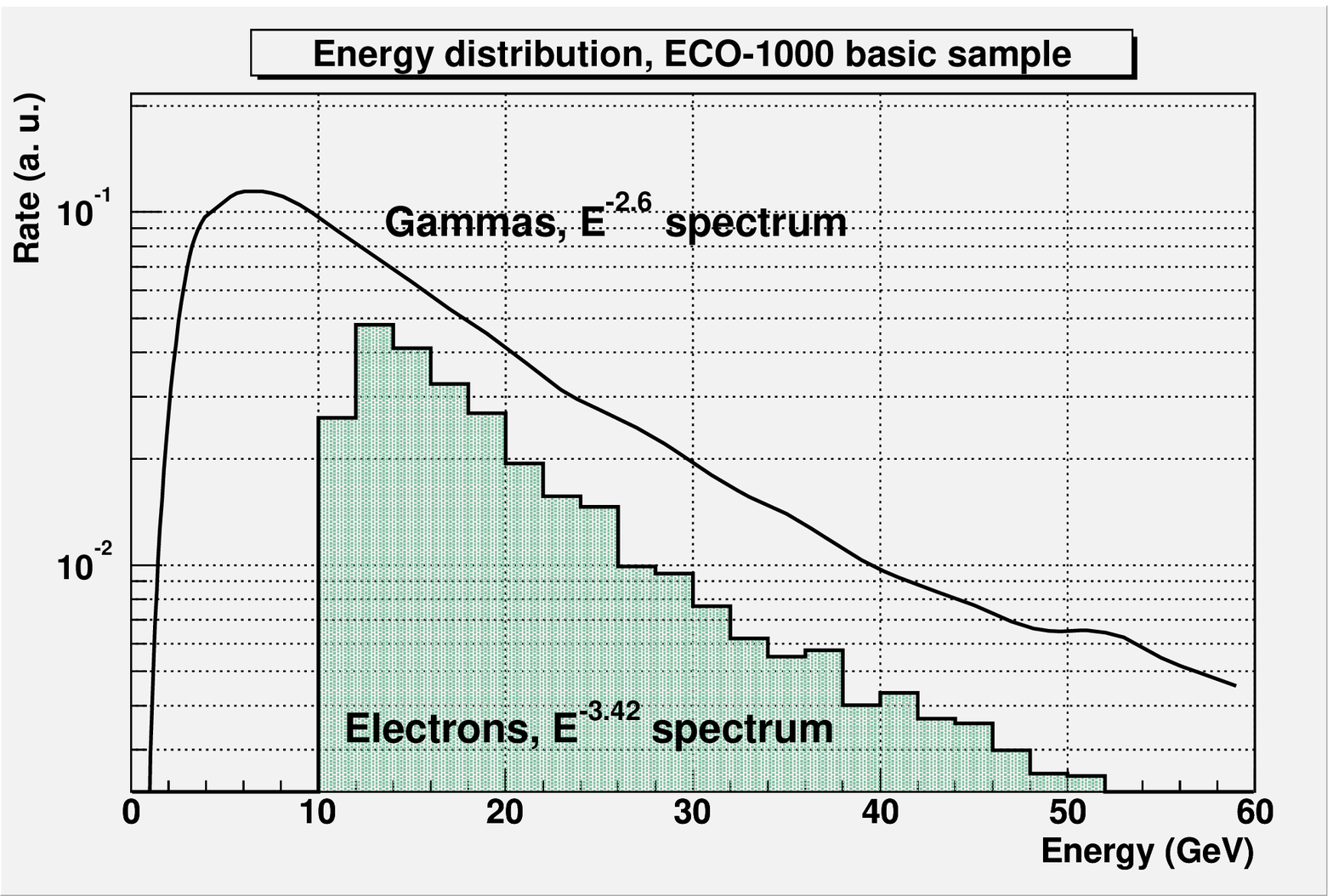}
\end {tabular}
\end{center}
\caption{\small \it Left: the direction-dependent cutoff rigidity 
for electrons above La Palma. Right: the sharp cut caused by 
inclusion of the geomagnetic field (Monte Carlo events); the upper curve
gives the gamma spectrum triggered in ECO-1000,
the lower curve corresponds to electrons with geomagnetic cutoff 
(arbitrary units).}
\label{figgeomag}
\end{figure}

At low energies, below 100~GeV, say, the background 
introduced by diffuse electrons \cite{aguilar} adds to the hadron background
(at higher energies, the rate falls rapidly);
being difficult to reduce, as electromagnetic showers have the 
same shape for electrons and gammas, we have looked into this background 
in some detail.
At such energies, one has to take into account the geomagnetic
cutoff \cite{cutoff}: the earth magnetic field deflects low-energy
particles before they can reach the atmosphere. 
The geomagnetic cutoff is dominated by the magnetic field within
few earth-radii; it does not depend on solar activity and can be 
assumed constant over extended periods of time. This cutoff can be
calculated analytically for a dipole field \cite{stoermer}; for our 
needs, we used a more precise tracing program \cite{prog}, 
containing a model of the geomagnetic field, and calculating for any position 
and incoming direction the
probability for a particle with given charge and rigidity\footnote{rigidity 
is defined by particle momentum / charge} to reach
the atmosphere. Compared to the energy resolution of IACTs,
this cutoff is very sharp (see figure \ref{figgeomag}).

We have estimated the electron rates in the telescope,
based on showers simulated in the energy range 2 to 200~GeV
for various zenith and azimuth angles, and using  
the energy distribution of electrons from \cite{aguilar}.
The geomagnetic cutoff and standard analysis
significantly reduce the expected electron background, 
the residual rates after trigger and after analysis are 
much below proton rates, see table \ref{tab:rates} below.

The action of the geomagnetic field on {\it showers} also is one of the 
factors limiting the performance 
of IACTs at low energies: the Lorentz force acts on the charged particles
in electromagnetic showers. They get deflected, mostly in an East-West 
direction, and some get trapped in the atmosphere. 
We have estimated 
this influence using Monte Carlo data for showers arriving 
under different azimuthal directions,
basing ourselves on geomagnetic field maps as available for La Palma.
We also have looked at the effect of the 
geomagnetic field on the effective collection area.
The conclusion is that the effect is quite visible, and will
require correcting flux measurements, at $\gamma$ energies below 100~GeV 
and for zenith angles 30$^\circ$ and larger. Possibilities to correct also 
the image parameters for individual events are under study; it is likely, however, 
that the classical way of analyzing events via few image parameters will
have to be substantially revised at low energies, independent of the field 
(see also section \ref{sec:analysis}).
\vspace{2mm}  \par
{\bf Isolated Muon Background}

\begin{figure}[t]
  \begin{center}
  \includegraphics[width=8cm]{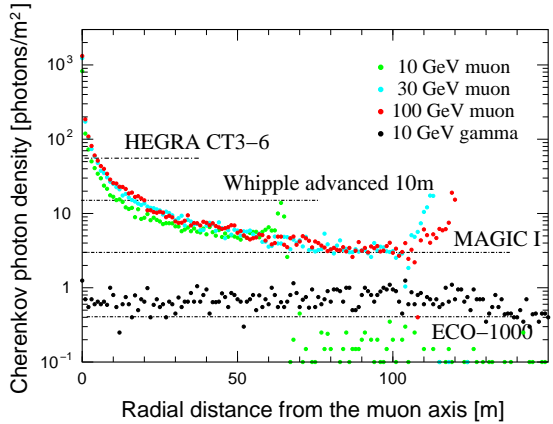} 
  \end{center}
  \caption{\small \it The lateral distributions of the Cherenkov 
photon density from single muons, for different particle energies.
Green, blue, and red dots denote 10~GeV, 30~GeV and 100~GeV muons, 
respectively; for comparison, 10~GeV gammas are also given (black dots)}
  \label{fig:lateral}
\end{figure}

Muons originating inside hadron showers can constitute a source of 
background. Muons travel a long path and often are above the threshold 
of Cherenkov radiation, a potential problem in case of high-energy showers.
At lower energies, muons are not a dominating background: for once,
lower energy primary particles generate only a small number
of muons, and secondly, muons generated from lower energy
primary particles such as $\leq$50~GeV, are less energetic themselves, 
and are mostly below threshold. 

Figure \ref{fig:lateral} shows the lateral Cherenkov photon density at 2200m a.s.l.
as function of the impact parameter and for different muon energies. There is
a rather sharp cutoff in the impact parameter
distribution for low energy muons: the angle 
between the Cherenkov photons and the muon trajectory is almost constant 
and small, $\sim 1^{\circ}$.
The ECO-1000 telescope (like MAGIC) can see the single muon up to the 
end of this cutoff. At higher energies,
the Cherenkov angle is larger and the collection area for muons increases.

\begin{figure}[t]
  \begin{center}
\begin{tabular}  {c c}
  \includegraphics[height=4cm]{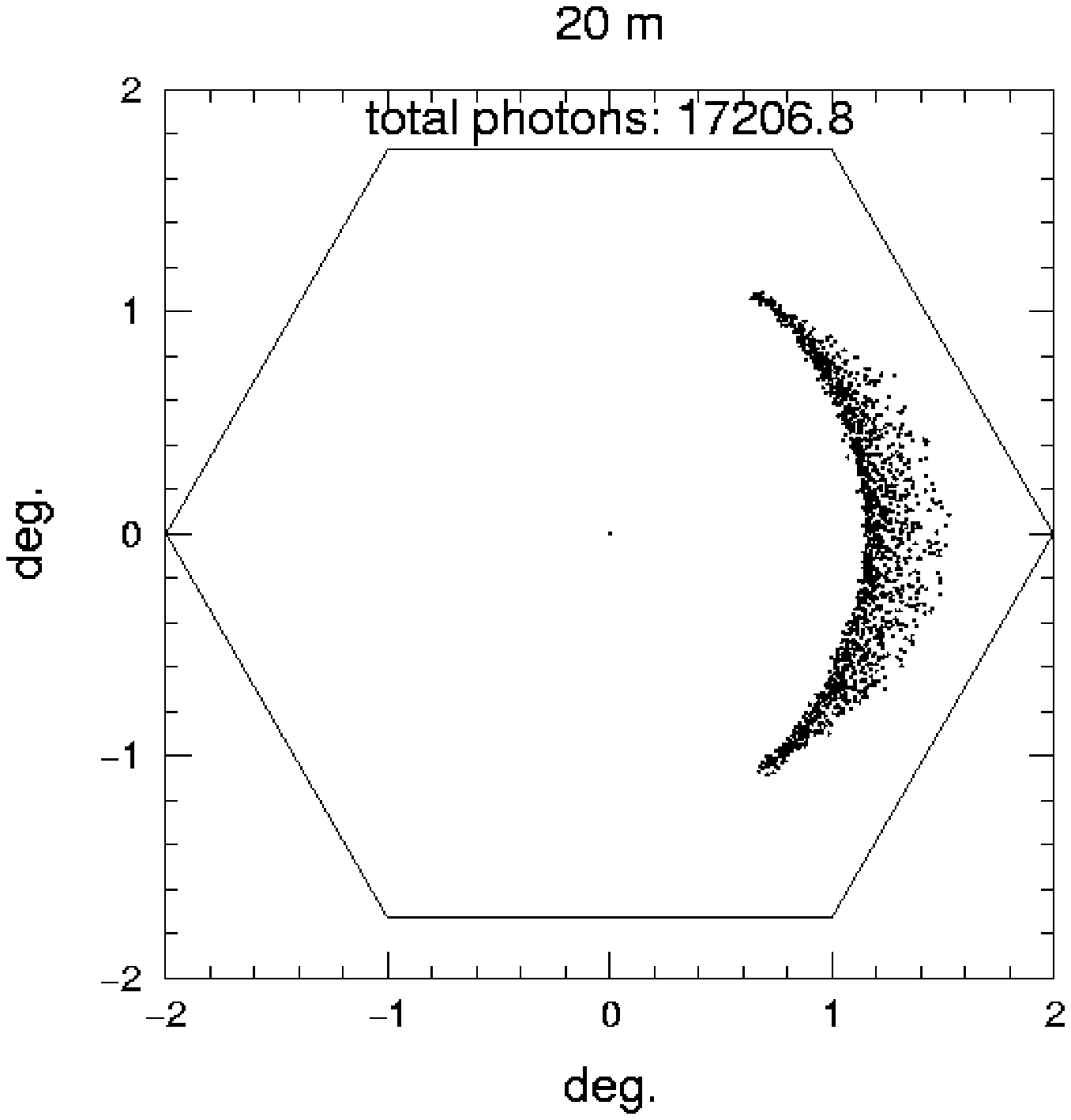}  &
  \includegraphics[height=4cm]{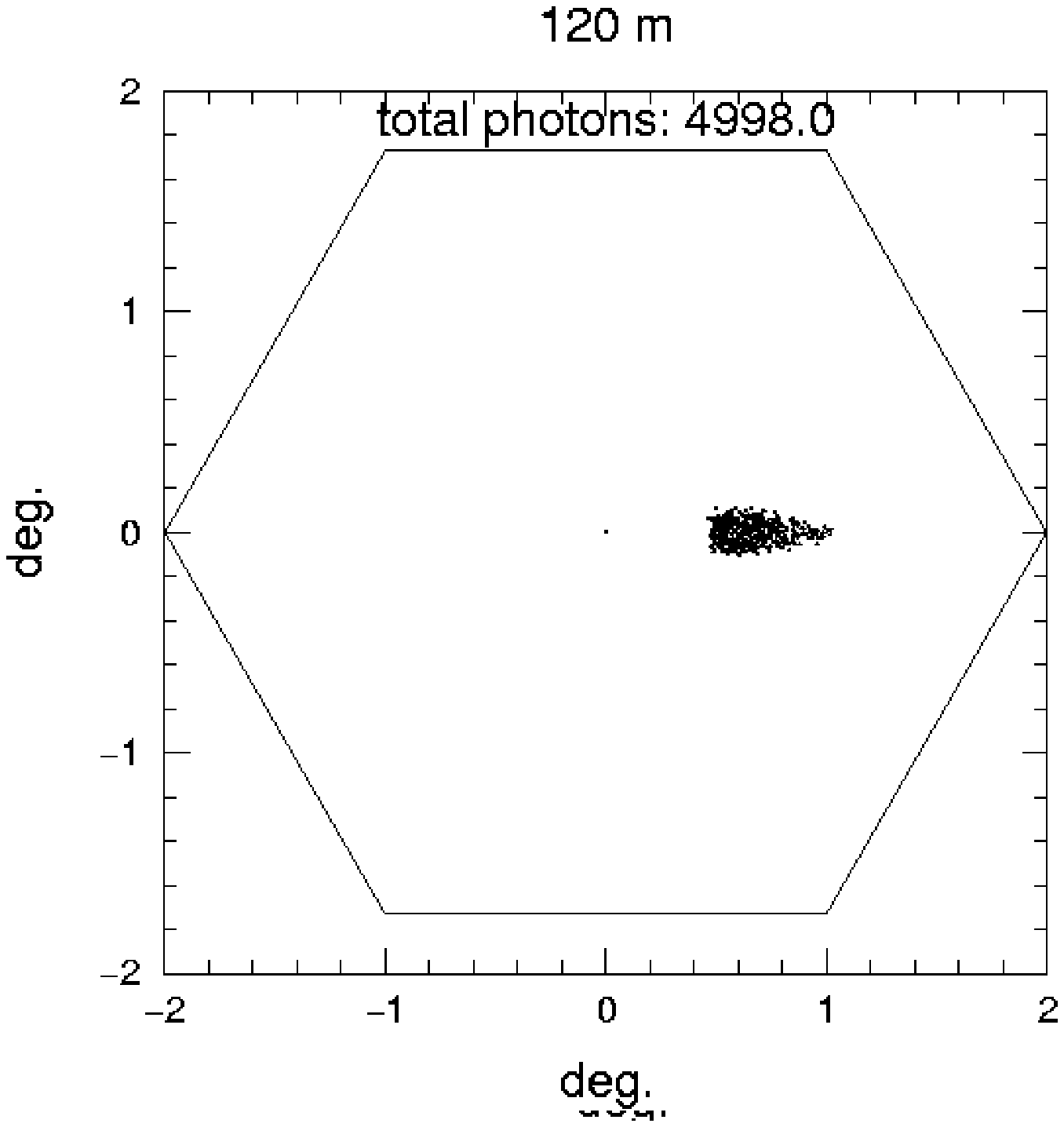}  
\end{tabular} 
  \end{center}
  
  \caption{\small \it Simulated single muon images in the camera plane. 
The images of muons impinging parallel to the telescope axis 
are shown for impact parameters 20~m and 120~m. 
The average total photon count is also indicated}
\label{fig:imagemase}
\end{figure}

A muon image in a Cherenkov telescope depends only weakly on the energy. 
For lower energy particles such as $\sim$10~GeV,
a muon image is typically brighter than a gamma image, if the particle is close
to the telescope; however, muon and gamma images look quite
different, they are easily distinguishable in most cases.
Single muon images on the camera plane corresponding to different impact 
parameters are shown in figure \ref{fig:imagemase},
for muons parallel to the telescope axis.
A full or partial ring can be seen up to an impact 
parameter $\sim$ 40~m. Typically, MAGIC or ECO-1000 will either not trigger on
such tracks, or they are easily discarded by image analysis.
For the images with impact parameters $\geq$ 40~m, images 
become increasingly more similar to those of gammas. The muon images
are elongated towards the camera center (for muons parallel to the telescope axis).
Muons travel over a large distance at high altitude,
and the change of Cherenkov angle is substantial; the length of the image 
(i.e. the width of the muon ring)
reflects this variation of Cherenkov angle. Some images of muon events 
do indeed resemble gammas, for large impact
parameters, as shown in figure \ref{fig:imagemase}.
Such events might be able to pass gamma selection criteria and add to the background, 
somewhat deteriorating the sensitivity of the telescope. 

\begin{figure}[t]
  \begin{center}
  \includegraphics[width=10cm]{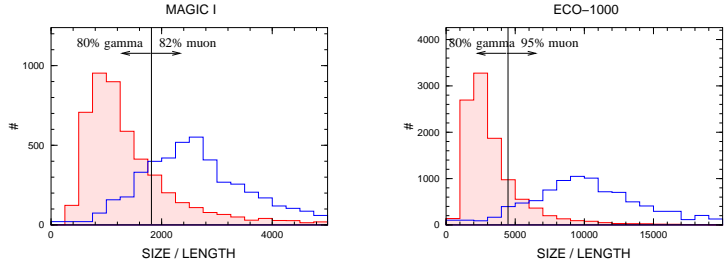} 
  \end{center}
  
  \caption{\small \it Distributions of the ratio size/length for 
MAGIC and ECO-1000. Red (hatched histogram, left) and blue (right) 
distributions denote gamma and muons respectively}
\label{fig:sizedivlen}
\end{figure}

In the analysis, we have additional handles to eliminate muons: 
we can use, for instance, the ratio of the image parameters {\it size} 
and {\it length}; as Cherenkov light is emitted 
isotropically in azimuth, and the radiated light per length is almost constant, this
parameter indicates that muons, on average, radiate over a longer distance, and it allows 
good discrimination, as seen in
figure \ref{fig:sizedivlen}. A cut which retains 80 \% of the gammas,
removes 95 \% of muons for ECO-1000, and  82 \% for MAGIC.
Further, the total photon count of muon events is higher than that  of
low-energy gammas ($\leq 30$~GeV), an additional parameter allowing better
$\gamma/\mu$ separation.

\vspace{2mm}  \par
{\bf Background Light} 
 
The background light of the night sky (LONS) is another factor in
analyzing image data from air Cherenkov  telescopes.  
In measurements of the Cherenkov photons from air showers, 
one invariably also integrates LONS,
because of the overlapping spectral distributions of these 
two different light sources. 
Although Cherenkov flashes are extremely short, and signal integration extends
over few nanoseconds,  the high intensity of LONS and the relatively large size  
of the imaging camera pixels (with a typical aperture of 0.10-0.25$^\circ$) 
result in some unavoidable noise component from LONS in the signal. A large 
telescope mirror, as foreseen for ECO-1000, will require increasing the trigger
threshold by a factor $\sim 2$, in stand-alone operation,
to avoid a high trigger rate due to LONS alone. On the other hand, 
for operation of several
telescopes in coincidence, the trigger threshold can be substantially
reduced, although the behaviour at low energies is subject to more fluctuations, 
and has yet to be fully understood.

The main contributions to the LONS are the airglow (excitation of air  
molecules high in the atmosphere during the day time and slow  
de-excitation during the night), the direct starlight, the integrated 
starlight, the zodiacal light (scattered sunlight in the ecliptic, 
depending on seasonal factors and  
latitude) and the aurorae (latitude and solar wind dependent).  
For sources in our galaxy, 
the light emission from the Milky Way makes the LONS several times  
higher than the LONS at high  galactic latitudes.      

\vspace{2mm}  \par
{\bf Background Rates} 

The above discussion has concentrated on background properties; it remains to
show the expected rates of the various background components.

The most important background is that of hadronic showers initiated by cosmic rays. 
We can define {\it muon events} as events whose ratio of the muon-induced
light in the camera to the total collected light
is larger than 0.1. 
If we follow the overall rates as in \cite{alcaraz}, we find from our simulations
the rates as in table \ref{tab:rates}. The attributes 
'triggered' and 'analyzed' refer to crudely
selected events and to events sent through standard analysis with gamma/hadron 
separation, as done for high-energy gammas. As will be argued in section 
\ref{sec:analysis}, these are undoubtedly pessimistic rates.

\begin{table} [t]
\begin{center}
\begin{tabular}{c c c c c c c}
   & all protons & all protons & $\mu$-events & $\mu$-events  & electrons & electrons\\
   & triggered   & analyzed    & triggered     & analyzed   & triggered   & analyzed\\
MAGIC        & 302  & 7.2    &  155  &  2.6  & 3.95  & 0.52 \\
MAGIC-HQE & 313  & 13.1   &  155  &  3.9  & 5.25  & 0.69 \\
ECO-1000     & 820  & 144    &  290  & 16.6  & 36.9  & 12.9 \\
\end{tabular}
\end{center}
\caption{\small \it The expected rate per second for contaminating proton showers,
for proton showers with muon content, and for electrons. 
More explanations are given in the text.}
\label{tab:rates}
\end{table}

\begin{figure}[t]
  \begin{center}
  \includegraphics[height=6cm]{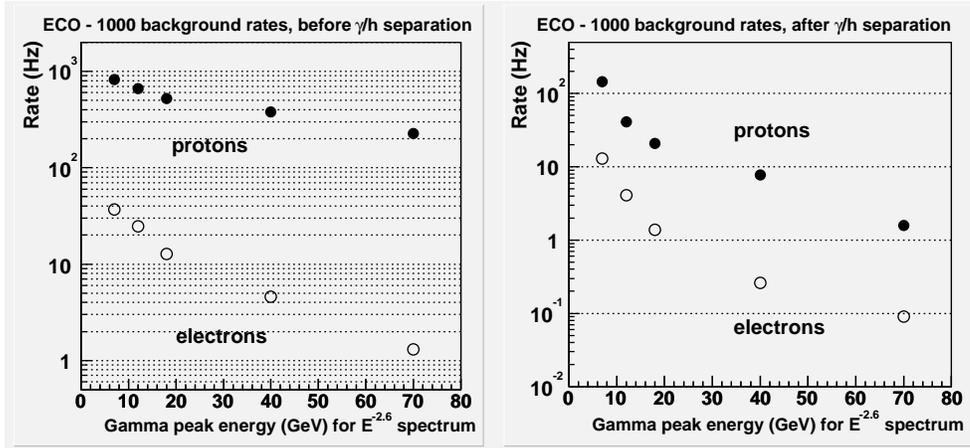} 
  \end{center}
  \caption{\small \it The figure shows the integrated 
proton and electron rates, at trigger (left) and analysis level (right), 
in the ECO-1000 telescope. The event samples were
selected by the minimal total number of analyzed photons in the image. This
selection clearly purifies the gamma sample, but also shifts its energy 
peak, viz. the energy threshold, shown on the x-axis.}
\label{fig:bkgcut}
\end{figure}

One notices that the total hadron background for ECO-1000 is about 
three times higher than for MAGIC: this is primarily a consequence of
the lower energy threshold, as can be seen from 
figure \ref{fig:bkgcut}. We refer again to the discussion in section \ref{sec:analysis}.
For muon events, the ratio is smaller, for
electrons the rate is negligible. He nuclei add some 15~\% to the proton rates.

For the background light (LONS), we have carried out 
measurements at La Palma \cite{lons}, 
which allow to conclude\footnote{the LONS intensity measured in a wide-angle pixel 
is clearly higher, after normalization for
the field of view: this is related to an additional contribution from 
bright stars, more likely as the viewed part of the sky increases;
our measurements showed that the normalized LONS intensity integrated in one steradian  
is almost twice as high compared to that measured in an angle  
$\leq 1^{\circ}$}, for the interesting wavelength 
range of 300-600nm, that the normalized LONS can be described by
$\sim 2 \times 10^{12}$ photons / (srad sec m$^2$).

With this intensity and the parameters of MAGIC, we estimate
that LONS induces on average $\sim 0.16$ photoelectrons per nanosecond 
and per pixel in the imaging camera. Note that the PMT pulse  
integration time for MAGIC at the output of the receiver board is  
$\sim 2.5$ns ($FWHM$) and the resolution time for the trigger (gate) 
is 4-5 ns. The light yield for ECO-1000 will be four times higher 
(for the ratio of mirror areas), somewhat reduced for a shorter integration time; 
this assumes that pixels of the same angular coverage are used as in MAGIC.

\section{\label{sec3}Optimizing the Signal-to-Background Ratio: Conceptual Choices}

{\bf Choice of site}

We propose as site for the European Cherenkov Observatory 
the Roque de los Muchachos, La Palma, the southernmost European territory.  
We plan to develop the site starting with the current MAGIC telescope. 
An eventual configuration including two or three $17m\oslash$ MAGIC 
telescopes with ECO-1000, all equipped with high QE cameras, will be a 
powerful stereo system. The La Palma 
site (28.8$^\circ$ N, 17.8$^\circ$ W, 2200 m a.s.l.)  is in many aspects very 
favourable,  both for astrophysics questions and for infrastructure issues.
A list of arguments, certainly incomplete, is the following:

\begin{itemize}
\item  a large section of the deep sky is visible, without obscuration 
by the galactic plane with its high light background, i.e. 
the northern sky is better suited for example for AGN and GRB studies;
 \item the galactic center is still well visible with a large 
collection area, with the energy threshold close to 25~GeV (i.e. with ECO-1000);
 \item the best studied object acting as a standard candle, 
the Crab nebula, is well visible from La Palma. It should take less than 
a few minutes to carry out calibration measurements with  high significance 
($>$8$\sigma$);
 \item optical visibility and weather conditions on the La Palma 
site are among the best world-wide;
 \item the Canarian islands are the best suited place for an 
observatory on European grounds because of its most southern location in Europe;
 \item the small-size island La Palma has unprecedented  night 
time temperature stability, because its takes on average half an hour to 
exchange the daily warmed-up atmosphere by a very stable air mass  from the sea, 
i.e. the telescope is not affected by temperature changes during nights;
 \item the site has already a very good infrastructure developed 
both  for optical astronomy and for the current gamma astronomy installation. 
Wide area connection to the European data nets are installed;
 \item the traffic connection to the main centers in Europe is excellent, 
i.e., the site can be reached during daytime within a few hours, and observations 
can be started the same evening; 
 \item on-site prospects to carry out simultaneous gamma and 
optical observations (AGNs, GRBs…) are excellent, apart from the fact that MAGIC has
already a close partnership with one of the multiple optical telescopes on La Palma, 
the KVA telescope;
 \item the population and local authorities have demonstrated 
a very positive attitude towards 
the needs of astronomical observations. There exists a special 
law to suppress background light from towns at the sea level.
 \item an observation site in the Northern Hemisphere is a necessary complement to
the likely southern gamma-ray observatory, also in build-up. 
\end{itemize}
\par
We want to elaborate briefly on the argument of altitude, i.e.compare a high altitude 
installation (5000~m) \cite{fiveat5, gamlca} with one at medium altitude (2200~m). 
At high altitude, the telescope is closer to the shower maximum; 
the Cherenkov light pool is smaller in area, and the intensity 
of photons/m$^2$ is higher, i.e., the threshold for an equal size 
telescope should be lower. Being closer to the shower does pose optical imaging 
problems, though, due to the limited depth of field of a single mirror 'imager'. 
Therefore, the diameter of the mirror has to be kept significantly smaller than 
for an installation at 2200~m, where  the collection area is larger by nearly a 
factor of two.  A camera with high QE, red-extended hybrid PMTs will be 
a bonus, making up for the higher losses in case of observations 
at large zenith angles.  Our long experience of working with different types of 
photosensors, always in collaboration with industry,
should give our collaboration an excellent basis to install a novel type of camera 
on a short time scale. The cost of building a high-QE camera will be more than balanced 
by the additional cost for a high altitude installation. 
In addition, with our acquired capital of know-how (having 
built the world-wide largest IACT) 
and our possibilities to test the new technological improvement at the La Palma 
site, we will further cut costs and also accelerate developments in general. 
It should also be mentioned that the La Palma site has 
enough room to install a stereo system of several ECO-1000 telescopes;
in addition,  a site of $>$1km$^2$ with slightly worse background light conditions  
is available on the neighbouring island Teneriffe.
\vspace{2mm} \par
{\bf Multi-telescope configurations}

We have at present initiated the construction of a second telescope of the MAGIC class,
largely unchanged from what is already installed in La Palma. 
This will give us, on a short time frame, several important advantages:

\begin{itemize}
\item  substantial flexibility in using the two telescopes either in 
coincidence mode on the same source, or observing two different 
sources, e.g. during GRB alarms;
\item  twofold mirror area, 
improving our capabilities in observing a single source;
\item  additional information for events falling into the overlap 
area, resulting in improved parameters and background rejection, 
and thus higher-quality events useful for normalizing
the observations taken outside the overlap area;
\item  an improved duty cycle, maintaining physics capability 
also during maintenance periods or technical runs; 
\end{itemize}

At the time of introducing a second MAGIC, we will be able to count on a 
substantial experience
with the first telescope, such that its running-in period will be much 
shortened (as will, of course, be its construction time).

We hope that we can improve the overall performance of the European Cherenkov 
Observatory by including advanced technical developments in the second 
MAGIC-type telescope, as this note suggests. 
Eventually, the experience in building it will enable us to propose
a telescope of the ECO-1000 class, within the next two or three years.

\subsection{Expected Performances}

{\bf Energy Threshold,  Collection Area, and Sensitivity}

\begin{figure}[t]
\begin{center}
\begin{tabular}{c c}
\includegraphics*[width=6cm]{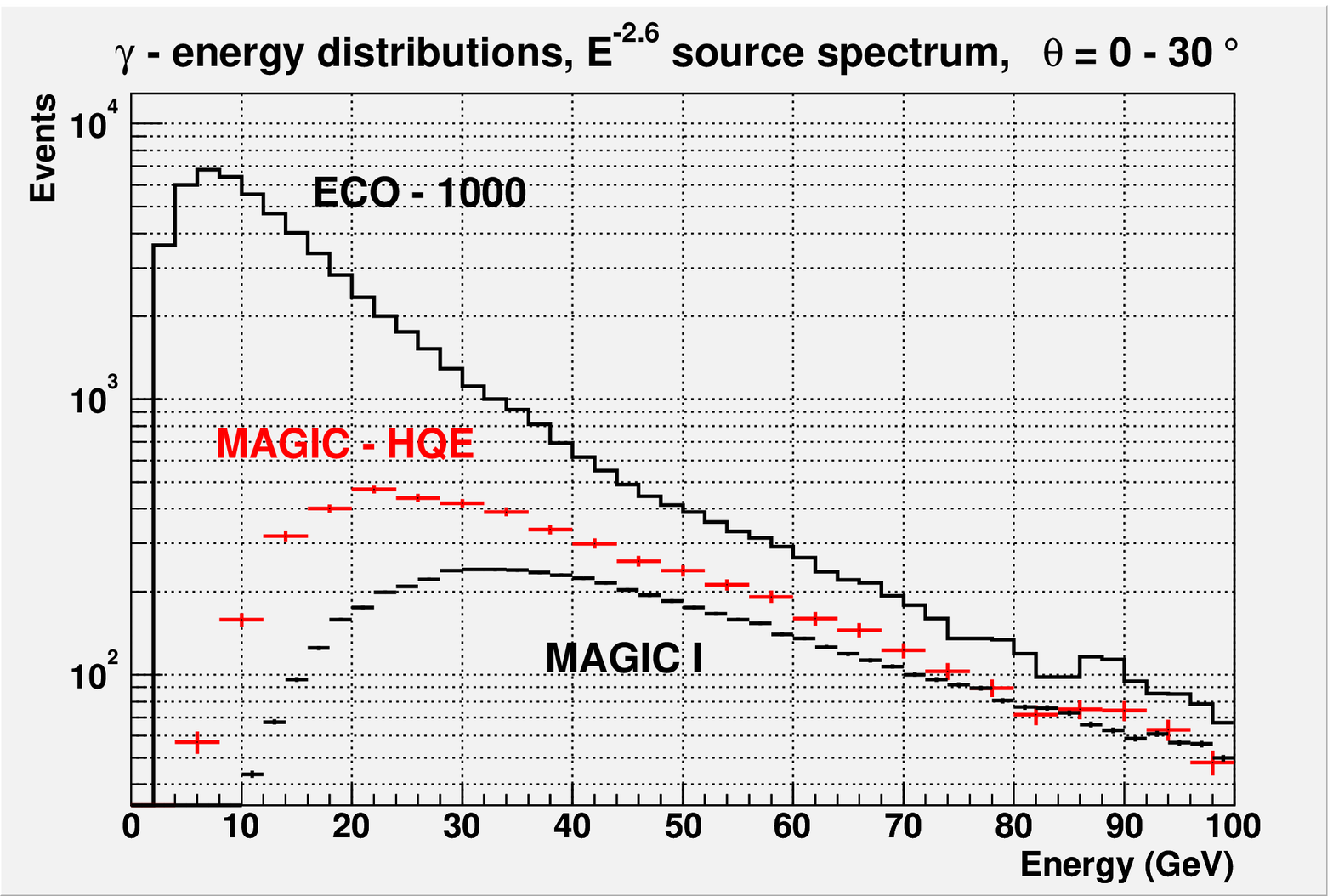}  & 
\includegraphics*[width=6cm]{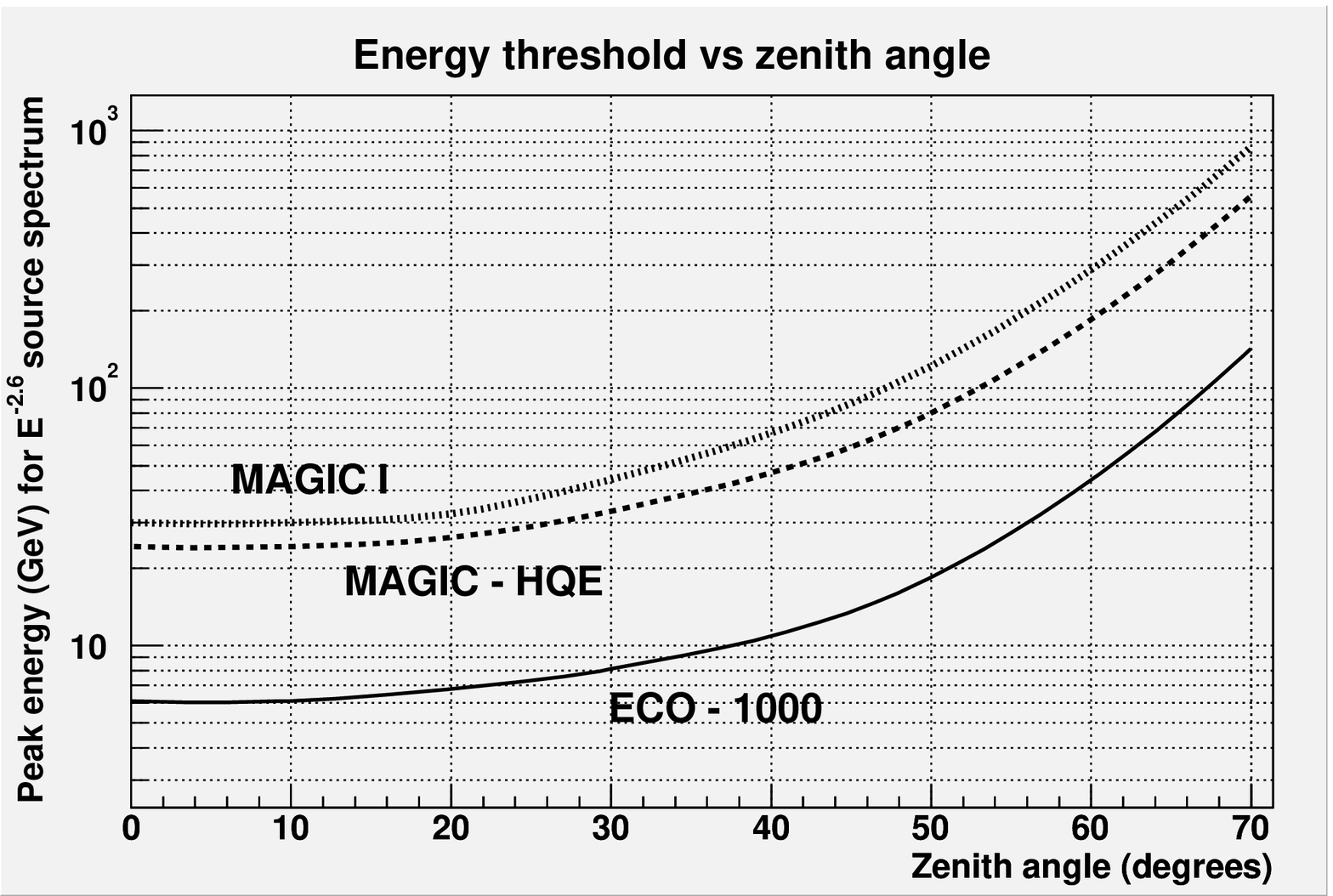}
\end {tabular}
\caption{\small \it Left: energy distributions 
at zenith angles up to $30^\circ$, 
for ECO-1000 (highest curve), for MAGIC-HQE (middle) 
and for MAGIC (lowest); right diagram: 
the energy threshold dependence on the zenith angle 
(for a different, loose trigger condition)}
\label{fig:thresh}
\end{center}
\end{figure}

Energy distributions after a conservative trigger ensuring acceptable data rates,
are shown in figure \ref{fig:thresh}, for ECO-1000, MAGIC-HQE, and MAGIC.
The energy threshold is defined to be at the 
maximum of the differential energy distribution. Changing the trigger conditions can lower this
maximum, but rates increase and the data quality suffers.  
The peaks shift somewhat after 
analysis, and are found at 7, 25, and 35~GeV for the three telescopes, 
respectively, for zenith angles up to $\approx 30^{\circ}$. 
Beyond, the threshold increases with the zenith 
angle; at $50^{\circ}$, it is about 19, 80 and
125~GeV for the three telescopes, respectively.

\begin{figure}[t]
\begin{center}
\includegraphics*[width=6cm]{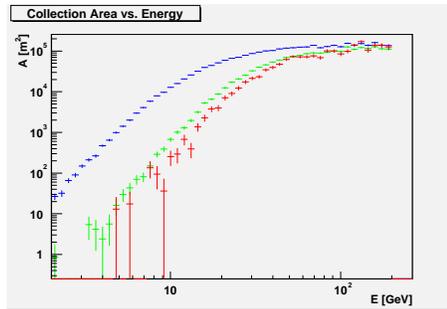}  
\caption{\small \it Collection areas for ECO-1000 (blue, leftmost)
MAGIC-HQE (green, middle), and for MAGIC (red, rightmost)}
\label{figcollar}
\end{center}
\end{figure} 

\begin{figure}[t]
\begin{center}
\begin{tabular} {c c}
\includegraphics*[height=5cm]{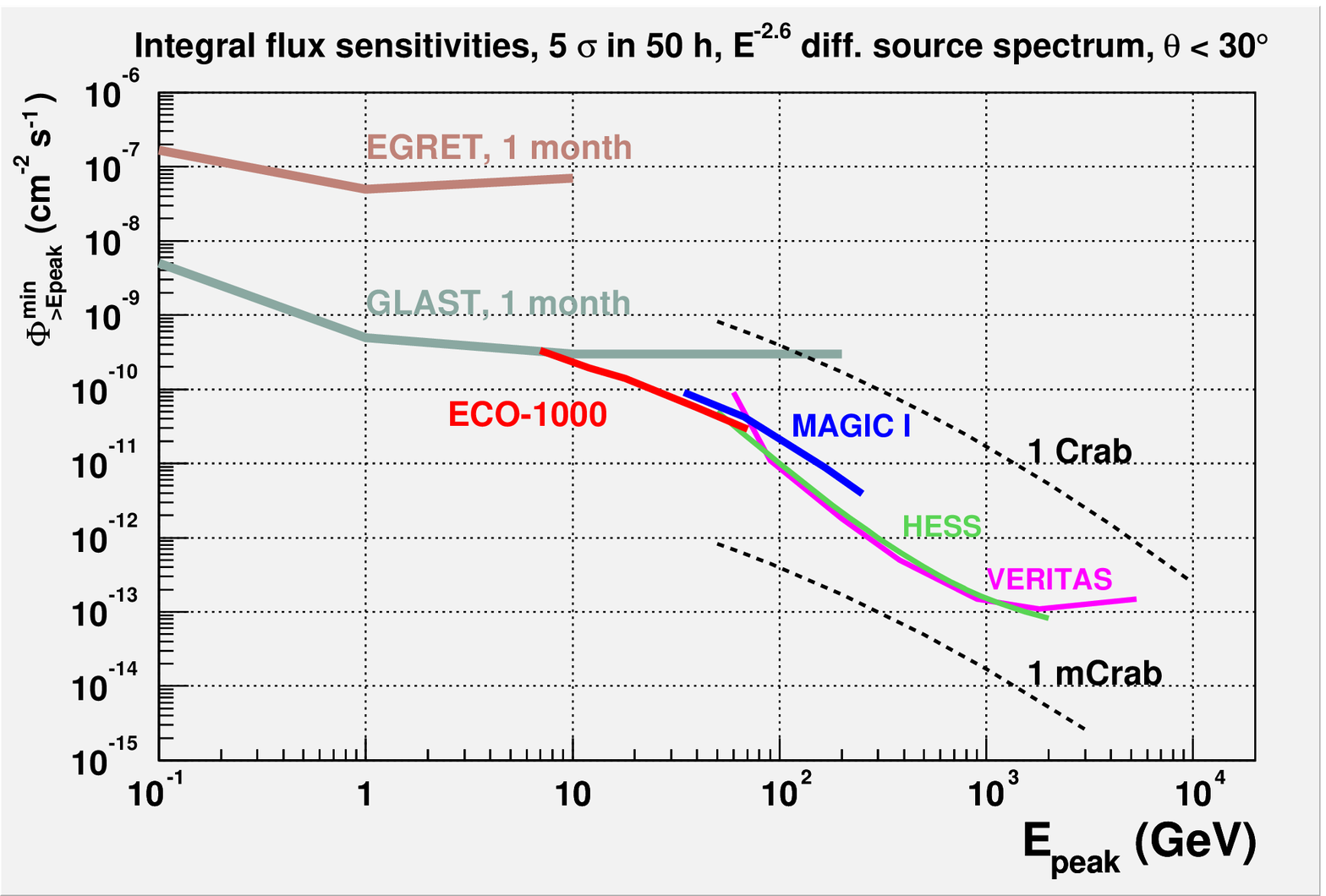}   &
\includegraphics*[height=5cm]{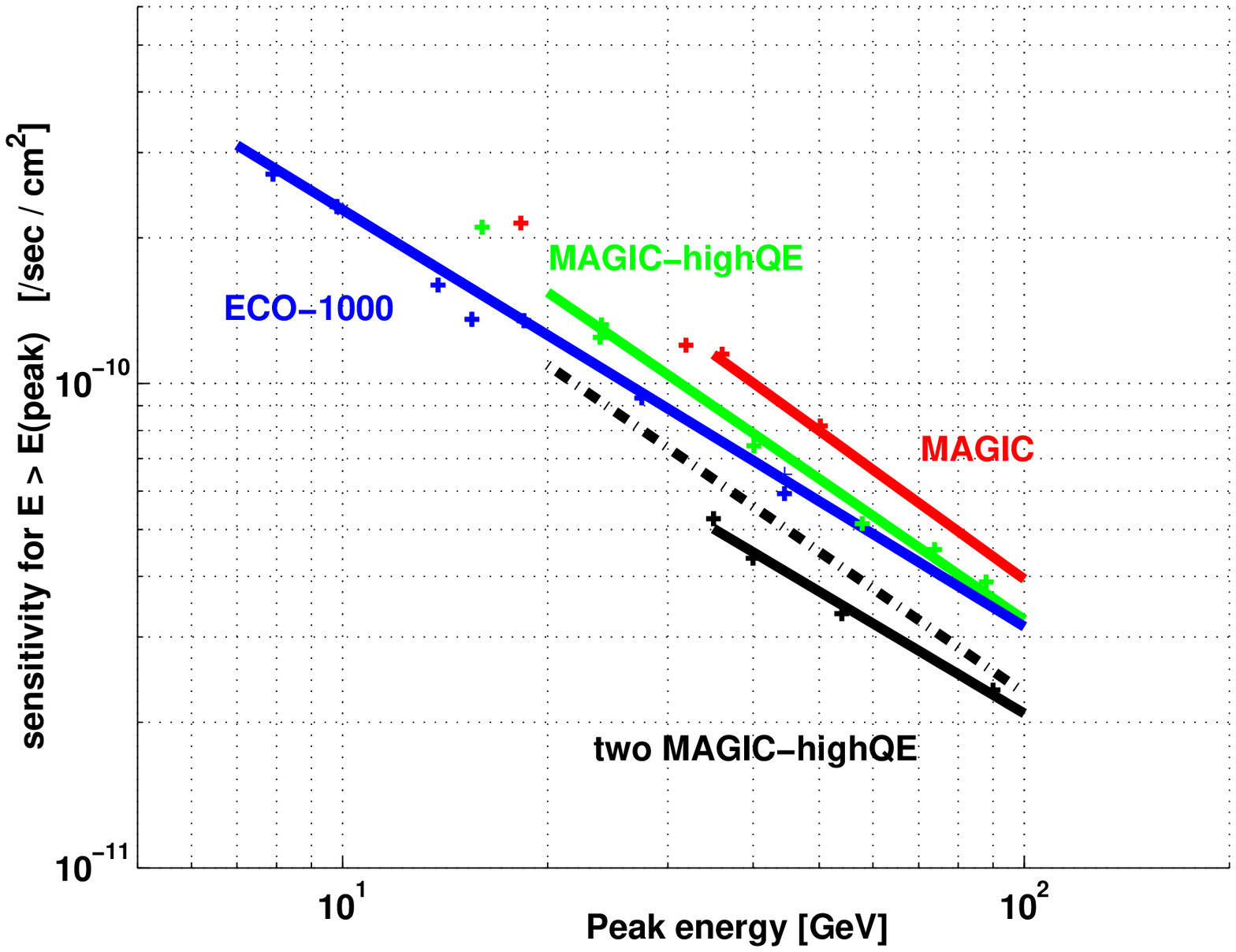}
\end{tabular}   
\caption{\small \it Left: comparative estimate for the sensitivities of MAGIC and
ECO-1000 with other instruments, existing or planned; the minimal flux for a $5 \sigma$
observation in 50 hours (one month for EGRET and GLAST) of observation 
in stand-alone mode is given. 
Only the low-energy region is shown for ECO-1000 and MAGIC.
Note that the curves for HESS and VERITAS correspond to multitelescope systems, 
4 telescopes for HESS, 7 for VERITAS.
Right: internal comparison at low energy for MAGIC,
MAGIC-HQE, and ECO-1000, plus two-telescope combinations; 
 upper points (red): present
MAGIC; middle points (green): MAGIC-HQE; lower points (blue):
ECO-1000, also with a high-QE camera; lowest (black) line: 
two MAGIC-HQE telescopes run in
coincidence; the dash-dotted line gives the value for two MAGIC-HQE telescopes 
run in stand-alone mode.
Conservative threshold energies are given by the start of the lines,
points  at energies below threshold correspond to events difficult to reconstruct; 
the power law line is only to guide the eye.}
\label{figsensit}
\end{center}
\end{figure}

Collection areas are shown in figure \ref{figcollar}; again, 
they reflect the different energy thresholds.    
Sensitivity estimates at energies below 100~GeV from a specific 
low-energy Monte Carlo simulation are shown in figure \ref{figsensit}. 
We see that at all energies the telescope sensitivity improves 
by introducing more efficient photosensors, and again by increasing the 
mirror surface, and the gains in energy threshold from MAGIC to MAGIC-HQE 
to ECO-1000 are substantial. Two telescopes run in coincidence (in this case at 85m 
distance) do not reach the low energy threshold; this is much in agreement 
with the discussion in section \ref{sec:analysis}, and clearly needs further study.
The proton background flux was assumed to follow \cite{alcaraz}, He ions were not
considered.

\section{\label{sec:analysis}Analysis of low-energy Gamma rays} 
\subsection {Gamma-Hadron Separation}

The separation of the signal events, caused by gamma ray showers 
in the atmosphere, from the background of cosmic rays, is a critical performance 
parameter for Cherenkov telescopes. The background, typically hadronic showers
initiated by protons or He ions, usually dominates the signal by at least two to three
orders of magnitude, depending on the source being observed. Much of this background
is eliminated by the fast trigger, implemented as hardware or firmware. In case of 
multiple telescopes, and if only the overlapping area is being used, already the 
coincidence trigger
can ensure a reasonably clean selection, at least at higher energies.
For stand-alone telescopes, the selection
must be based on the properties of a single recorded image, and is more of a challenge 
to the statistical analysis methods, although the final results are comparable 
\cite {Fegan, Kranich}.

In accelerator experiments, the separation of particles and energy estimation rely 
on calorimetric measurements
at higher energies, and use single-track information at lower energies,
where the statistical fluctuations make calorimeter information less reliable.
The ground-based Cherenkov technique does not have this alternative, so we 
have to make the best of the information left over by the showering process
in the atmosphere.
 
The Cherenkov light generated by electromagnetic showers at low energies, 
clearly below 100~GeV, can no longer be adequately approximated as a light pool
of uniform illumination arriving at the detector. This is basically the 
assumption made by the classical analysis methods
introduced by Whipple \cite{Fegan}. 
These methods rely on few parameters, to which the image information 
is reduced. Typically, the image of a shower, after some pre-processing, 
is an elongated cluster with its long axis oriented towards the camera center 
(assuming a point source
and a shower axis parallel to the telescope axis). A principal component analysis   
is then performed in the camera plane (also called second-moment analysis), 
to obtain characteristic image parameters. 
Subsequently, the methods use, explicitely or implicitely, the multi-dimensional 
space spanned by these parameters, relying on Monte Carlo data of gamma 
events to define gamma
properties and to discriminate them from cosmic ray (hadronic) showers. 
Although the basic image parameters introduced by Hillas and 
used in the image analysis at high energies, have been
amended many times, results are deteriorating dramatically as shower 
energies decrease \cite{ghpub}.

\begin{figure}[t]
\begin{center}
\begin{tabular} {c c}
\epsfig{file=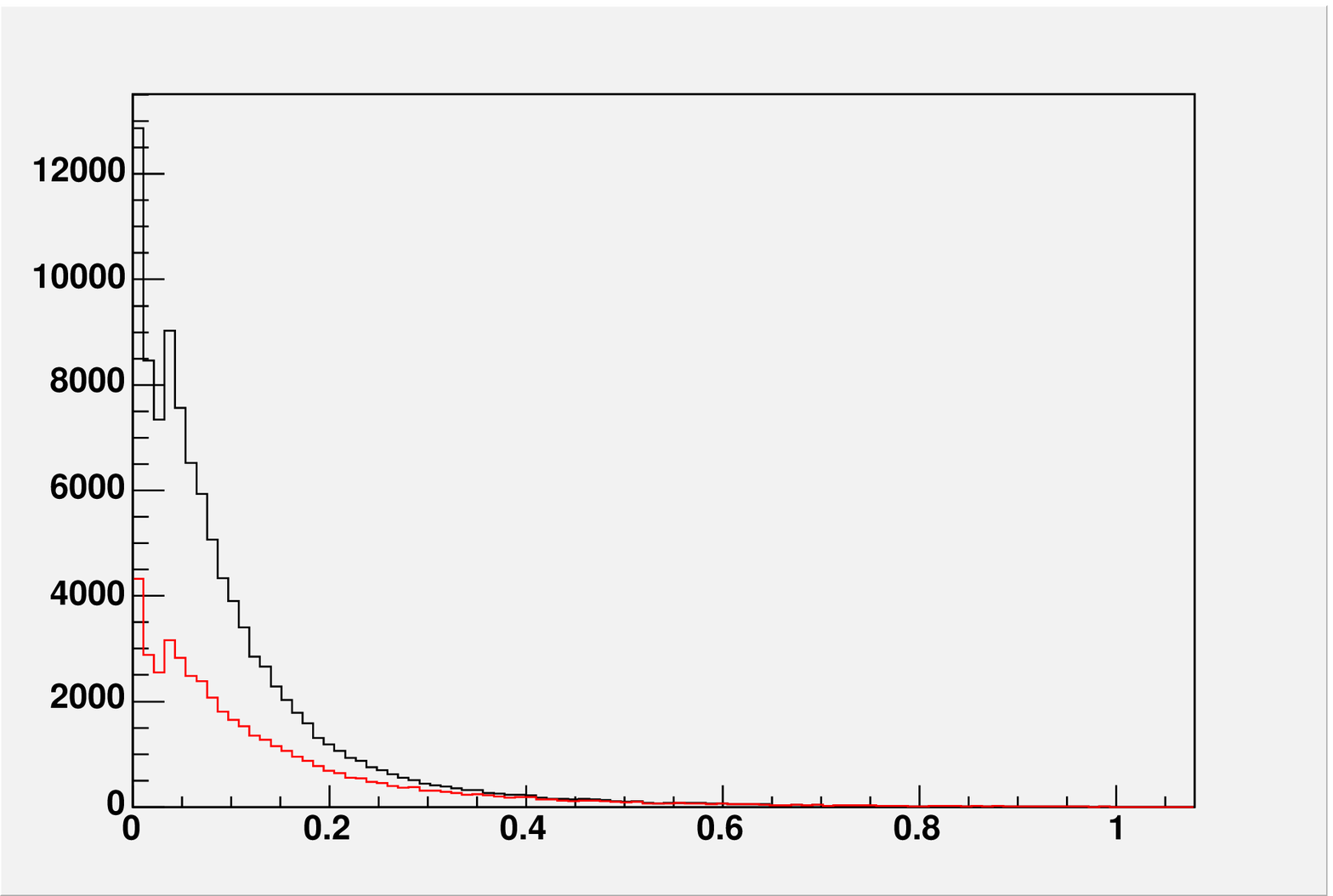,width=6cm} &
\epsfig{file=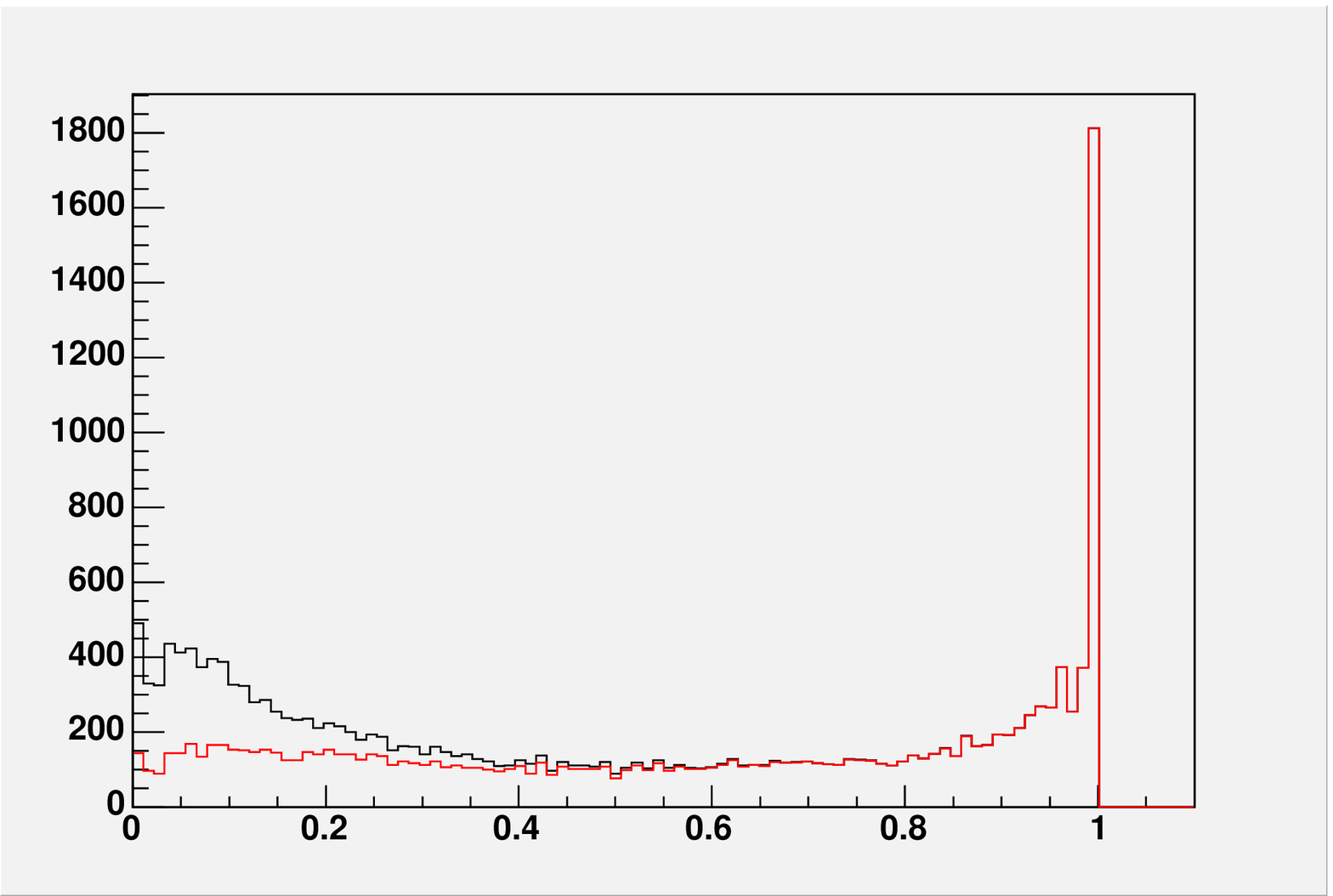,width=6cm}
\end{tabular}
\caption{\small \it The probability of events to be a hadron (hadronness
explained in the text, and shown on the x-axis)
is histogrammed for two different cuts in the total number of photons analyzed.
Left: gamma events of energy $<$100~GeV;
Right: hadronic background events remaining after trigger.}
\label{fig:hadronness}
\end{center}
\end{figure}

This should not be surprising: at low energy, the photons 
observed in the camera undergo large 
fluctuations in number, and in incident angle and position:  
the information is thinned out substantially compared
to higher energy events, which fill the light cone rather uniformly. 
At the same time, the overall probability of detection is reduced, 
and an increasing number of background hadronic showers have a 
tendency to look like gammas, due to single 
$\pi^{\circ}$-s. This is shown in the diagrams of figure \ref{fig:hadronness}. 
We plot there the {\it hadronness} for two different cuts in the total 
number of photons remaining 
in the analysis. Hadronness is a variable containing, in our analysis, all 
information used in gamma/hadron separation (we use a multi-dimensional classification
procedure called {\it Random Forest}, see \cite{ghpub}).
Cuts in the total number of photons are related to energy cuts; they do remove the 
gamma-like hadron background, but also scale down severely the 
observed gammas - and obviously influence the energy threshold. 
 
\begin{figure}[t]
\begin{center}
\epsfig{file=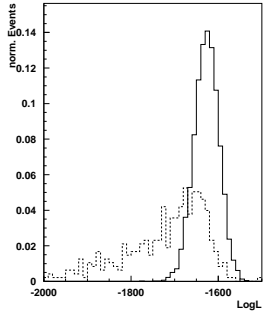,height=5cm} 
\caption{\small \it 
Log-likelihood values for MC-generated gamma (solid line) and hadron
(dashed line) showers, for the MAGIC telescope}
\label{figlogl}
\end{center}
\end{figure} 

Our studies with low-energy events make us believe that the original method with
few image parameters, 
developed for events with hundreds of GeV of energy, will seriously break 
down at energies below 50~GeV, requiring novel analysis methods. On a sample of pure
Monte Carlo events in MAGIC, we have noted, using the classical separation method,
reduction factors for hadronic showers
(after triggering and pre-selecting events) going from $\sim 400$ for 
events above 120~GeV to $\sim 10$
for events below 60~GeV \cite{ghpub}.
Some of this effect is compensated by the smaller number
of triggered hadronic background events; 
we are also confident that the limits of analysis can be
pushed further than presently known. The low-energy data now coming from MAGIC,
will be of invaluable help in adjusting our analysis to this new energy
range in the future. 

We are working along several avenues, the most obvious 
being to include in the analysis some
information so far untapped, like arrival time. 
In an alternative to the pure second-moment analysis, we include multiple 
new parameters, e.g.
a measure of the 'lumpiness' of events: in general, the light from 
a gamma-ray shower is closely grouped together in the camera, while the
light from a hadron shower is spread more widely, in separate clusters. 
We attempt to use this by describing the shower image in the camera 
by a 2-dimensional function (multivariate Gauss or similar); 
the free parameters of this function
are derived from a maximum likelihood fit, using all pixels without
preceding image cleaning. A given
2-dimensional function of this type will better fit the gamma showers, and the
derived log-likelihood value can thus be used to distinguish between
gammas and hadrons. Figure \ref{figlogl} shows the distribution of 
this variable. The gamma sample used here
consists only of showers with energies below 40~GeV.

We are also interested in using established semi-analytical methods,
which fit directly physical shower parameters like energy, shower maximum, 
impact parameter etc. \cite{lebohec, demaurois}. They seem to perform well 
at higher energies, but still have to show how they deal with lower-energy showers.

So far, these studies are in an experimental stage: all analysis results shown 
in this note have been obtained using the classical
image parameter method. At low energies, these results can clearly be considered to be 
pessimistic.

\subsection {Energy Resolution}

For higher energies, we have estimated the energy resolution using different methods, 
all based on various expressions involving calculated image parameters.
All methods show a relative energy resolution of between 25 and 30~\% for events
between 100 and 200~GeV. 
At lower energies, the same methods can not be used without changes. 
The problems are due both to limits in the 
physics\footnote{low-energy showers fluctuate dramatically both in space and in 
the fraction observable through Cherenkov radiation}
and to biases in the reconstructed event samples. 
Even the meaning of 'resolution' needs defining at low energy, as the statistical
distributions are wide and skew, viz. highly non-Gaussian.

This work is still in progress, and subject to the remarks given in section
\ref{sec:analysis}. 
An estimation of energy without reconstruction, based
on the photons that hit the telescope, results in an energy resolution
of $\pm$50-60\% (defined as half of the width containing 68\% of the events),
for an incident energy of 10~GeV. More precise results
at low energy will be available as MAGIC data are fully understood.

\section{\label{sec4}The Practical Implementation}  
The goal of our study is to prepare and test the technology for an ultra-large 
IACT, dubbed ECO-1000. It can be operated either as stand-alone telescope or in 
coincidence mode, as part of the European Cherenkov 
Observatory. The technology tests and studies we propose will pave the way for a 
rapid realization of such a telescope 
and will allow to make it available to the community on a predictable time scale. 
Current analyses have shown that within the next decade no viable alternative
to the proven concept of IACTs will be available, in performance, technology and costs,  
for ground-based gamma astronomy.

The general technical concept is basically an extrapolation from existing, 
smaller-diameter IACTs; several new solutions have to be found, however, 
to cope with the much more 
demanding operating conditions of a large telescope and to 
achieve a lower energy threshold. 
Our experience in the past has been that the change from a 10 m class IACT 
to the 17 m MAGIC telescope did require several design changes, 
the most obvious one being the introduction of the active 
mirror control, to overcome the support frame distortions. The envisaged linear 
factor two in size when going from MAGIC to ECO-1000, can be expected to 
again set new demands. The present note, suggesting design and feasibility studies,
along with some prototyping, addresses the areas in which such preparatory work
seems most relevant.

\subsection{\label{sec41}Main technical goals for ECO-1000}
They can be summarized as follows:
\begin{itemize}
\item  construction of a mirror with 34m diameter
 \item  increase of the System Quantum Efficiency over 
present numbers by a factor $\geq 2.5$   
 \item  improved mirrors with permanent active mirror control
 \item data handling capability up to 15 kHz trigger rate, and 
pulse height digitization with a sampling rate of 2.5 GHz (assuming the photosensor
are sufficiently fast), with a 10 to 12 bit dynamic range
 \item  repositioning time to any point on the visible sky within
$\leq 15$ seconds, for Gamma Ray Burst studies      
\end{itemize}
\vspace {3mm}
Additional goals we want to pursue are the following:

\begin{itemize}
\item  operation up to $100^\circ$ zenith angle
 \item  field of view sufficient to cover shower images from extended 
 sources  of $0.5^\circ $ to $1.0^\circ$ diameter, viz. $\geq 4-5^\circ $ in diameter
 \item  operation also during periods of moonshine, to improve the duty cycle 
 \item  tracking precision of $0.005^\circ$ 
 \item  power consumption of $< 40kW$  
 \item  price tag of $\leq $20~MEuro   
 \item  construction time $\leq 3$ years
\end{itemize}

It is most important to concentrate on studies and prototype implementation
of several key elements:
\begin{itemize}
\item Extensive, long-term and large-scale tests of a new type of high-QE hybrid 
photosensor, and construction of a prototype camera
 \item Engineering design study of the mechanics for a $34 m\oslash$ 
 lightweight telescope construction
 \item Development of new lightweight, large all-aluminium 
mirrors of hexagonal shape and $1.3 m^2$ area
 \item Permanent active mirror control without interference 
with normal telescope operation
 \item Detailed studies for a new user-friendly observation 
strategy to provide access for external users, participation in multi-wavelength 
campaigns and rapid dissemination of the results.
\end{itemize}

\subsection{Highest priority: new high-QE hybrid photosensor}

Lowering the threshold considerably below that of contemporary IACTs, 
cannot be achieved by increasing the mirror area alone. 
Optical requirements limit the maximal mirror diameter to 25-35m, 
depending on the telescope altitude and zenith angle. 
Any further increase in detection efficiency has to be achieved by increasing 
the conversion efficiency, i.e., the system quantum efficiency (SQE), 
of the initial photons to measurable photoelectrons. 
For a camera based on PMTs, the SQE is a product of several components:
\vspace{2mm} \par
  $ SQE = \int [R_m \cdot (1-L_{foc}) \cdot (F_{pmt}+ F_{lc} \cdot R_{lc})
 \cdot QE \cdot C_{d1} \cdot (1-L_{d1})] d\lambda $
\vspace{2mm} \par
with
\par   $R_m = R_m(\lambda)$ = mirror reflectivity at wavelength $\lambda$ 
\par   $L_{foc}$ = Focussing losses: fraction of light not reaching photosensors 
(typically 10\%)
\par   $F_{pmt}$ = fraction of active area of photosensors (typically 30-60\%)
\par   $F_{lc}$ = fraction of area of the light catchers $(F_{pmt} + F_{lc} < 1)$
\par   $R_{lc}$ = reflectivity of light catcher (typically 85\%)
\par   $QE$ = quantum efficiency of photocathodes
\par   $C_{d1}$ = photoelectron collection efficiency onto the first dynode
\par   $L_{d1}$ = losses at the first dynode including backscatter and fluctuations 
in the number of secondary electrons (10-30\%)

For a state of the art IACT, the SQE for wavelengths
between 300 and 550nm is at most 10-15\%. 
The SQE is dominated by the low QE of the PMTs, or by the product 
$QE \cdot C_{d1} \cdot (1-L_{d1})$.
Given the limitations of the mirror size, the main challenge then is to improve 
the photosensors. In close collaboration between the MPI-Munich and Hamamatsu 
the company is carrying out a development of 
new hybrid PMTs with a high QE photocathode and an avalanche diode as 
electron-bombarded anode with internal gain. The cathode material, GaAsP, 
has a QE of 45-50\% between 400 and 700 nm. Below 400 nm the QE drops smoothly to 
12\% at 300 nm. The use of a Silicon avalanche diode, an 8 kV cathode-to-anode 
voltage and an avalanche gain of 30 helps to overcome most of the losses in 
classical PMTs due to a nearly 100\% collection efficiency, low backscatter and 
no additional amplification losses, The extended spectral sensitivity of
hybrid PMTs is an additional bonus, because the atmospheric effects distort 
the original $1/\lambda^2$ shape of the Cherenkov spectrum (see figure \ref{fig:ds4}a).  

Smaller size prototypes are now under test and have confirmed the predictions. 
First measurements indicate that the gain in overall efficiency compared to a 
classical PMT is between 2.5 and 3.5 (zenith angle dependent). 
Part of the gain is due to the extended spectral sensitivity. 
The use of such a type of photosensor should result in a major improvement 
of the SQE and, in turn, a lower energy threshold and higher sensitivity. 
As we deal with a new device, extensive large-scale and long-term tests 
under realistic conditions are needed. We propose to carry out such a test by 
replacing part of the intended classical camera for the next MAGIC-type telescope. 
It is justified to 
base such a test on about 15\% of the number of PMTs needed for ECO-1000, 
i.e. 450 units. The test camera would be partially equipped with classical 
PMTs and partially with the new hybrid PMTs, in order to have a direct comparison.

\subsection{Design study of the telescope mechanics}
Based on the excellent experience made with the MAGIC construction, 
we plan to use again a space frame based on carbon fiber-epoxy tubes.  
The advantages are a reasonably stiff construction, low weight and minimal thermal 
expansion, as well as excellent oscillation damping. The use of CF-tubes 
increased the price for MAGIC by about 10\%. For ECO-1000 a higher strength fiber 
is envisaged; industry is quite interested in such a development, and judges there
exists quite some market potential.

It will be necessary to generate a detailed computer model of the telescope structure 
and mirror support, and to simulate numerous features:

\begin{itemize}
\item Stability
 \item Deformations at different zenith angles
 \item Oscillatory behaviour
 \item Resistance against strong storms, ice loads etc.
 \item Thermal expansion
 \item Minimization of weight
 \item Blueprints for the construction
 \item Requirements for the foundation
 \item Possible air turbulences, which might affect 
nearby optical telescopes
\end{itemize}

We may resort to an industrial offer to carry out the entire study. 
Several design parameters depend on the telescope support details; this
study, therefore, is one of the most urgent items to tackle.

\subsection{Design and feasibility study of mirror elements, prototyping}

Our final goal is to build ECO-1000 with hexagonal mirror elements 
of $1.3~m^2$using a novel light-weight
construction. We want to field-test the production mechanism by producing a limited
number of mirror elements with the same method,
and incorporate them in a MAGIC-type telescope. They will be square $1m^2$, 
and will be produced as a single honeycomb sandwich 
panel, to which a native spherical shape is given during the gluing process.

The gluing procedure needs, therefore, a spherical mold on which a 8cm Al 
honeycomb is sandwiched between two 2mm Al plates (skins), and forced with a 
vacuum bag to the spherical surface of the mold. The gluing itself will 
subsequently take place in a pressurized oven, at high temperature (150 deg), 
with a cycle of pressure/temperature typical for aero-space qualified 
gluing procedures.

The additional cost of producing a native spherical shape is more than balanced by the 
savings in the milling of the raw spherical shape, both in the cost of the 
machining and the subsequent manual labor.

Using this technology, the thickness of the front plate does not need to 
be very large because it has already the right spherical shape, so the 
mirror can be as light as 15 $kg/m^2$. This parameter is very important 
for the total weight budget for a large telescope.

Pre-shaping and polishing smaller mirrors has already been done successfully. 
Preparations for producing the first mirror of $1m^2$ surface
are under way, along with the tooling for the serial production.
This development will have to pass through several
iterations of prototyping and measurement; we estimate
a total prototype production of at least 100 elements.

\subsection{Design and feasibility study of the permanent 
active mirror control, prototyping}

Larger mirror diameters put exponentially growing demands on the 
stiffness and weight of the support frame, if required to be rigid. 
Alternatively, an active mirror control can be 
used and thus allow for the modest deformations of a much lighter structure. 
Such an approach follows the trend 
in modern optical astronomical telescopes, although the implementations and 
detailed requirements are quite different. As the main mirror has to be focused 
on an altitude of around 10-30km, where the showering occurs in the atmosphere,
it is not possible to use guide stars for 
the control. Instead, one can check the orientation of the segmented 
elements of the mirror by laser beams. For MAGIC, a system based on red lasers 
was used, thus requiring periodic interruption of the observations for mirror readjustments. 
For ECO-1000 it is planned to have a permanently active system 
based on infrared lasers, which do 
not interfere with the photosensors. Some basic development is needed, but the 
first suitable industrial components are available. We plan to design such
a system and implement it on a small scale. Installation on some of the
novel mirror elements in the next MAGIC-size telescope 
for long-term tests is foreseen.

\subsection {Studies to prepare the European Cherenkov Observatory (ECO)  
for access to a much widened community}

During the early phase of ground-based gamma-ray astronomy 
the experiments were designed and 
carried out similarly to high energy physics experiments at 
accelerators, i.e., as very specialized instruments 
to be frequently changed and operated for a rather limited 
duration by a dedicated collaboration. 
Also, analysis and physics interpretation could only be carried 
out by the collaboration. 
Currently, trends are to convert the detectors into facilities 
and to make them accessible to a much wider 
community. Here we mention in particular our intention to open 
the telescope facility to guest observers 
(not required to have detailed knowledge of the instrument 
features) and to integrate most of the physics 
program into multiwavelength observations. This requires a 
considerable change in structure and coordination of operation, 
data standardisation and means of fast data dissemination. 
An important issue will also be the availability of 
standardized Monte Carlo simulation packages, a 
necessary prerequisite for the extraction of the final results.
Here we propose to carry out such a study, define 
the necessary changes for the operation strategies and 
necessary hardware.

\subsection{Environmental Impact Studies}
 The installation of a new telescope requires studies of the environmental 
impact, similarly to those carried out for MAGIC.
The following details have to be studied:
\begin{itemize}
\item Geology of the ground 
 \item  Impact on the ecosphere
 \item    Impact on the ecosystem by the additional infrastructure
 \item  Telescope stability in case of strong storms and in case 
of ice deposits
 \item    Impact on possible historical remains
 \item    Impact from the power release
 \item  Impact from possible generation of atmospheric 
turbulences that might affect the nearby optical instruments. 
\end{itemize}
This study requires outside expertise and special tests, 
e.g. wind channel tests.

\section{Conclusion} 

Multiple strong arguments can be made for the extraordinary physics potential
that can be tapped by gamma ray telescopes with the lowest possible energy threshold.
A key instrument to aim for is a telescope with a large mirror surface and a high
quantum efficiency camera. We call this telescope ECO-1000, and have presented above
the necessary steps to work towards such a device over the next three years.
One important step is a design and feasibility study, including some prototyping;  
we propose here to equip part of a MAGIC-size telescope, now under construction, 
with novel photodetectors, with optimized mirrors, and with 
a permanently active mirror control system. Use in a working telescope
will not only allow long-term field testing of these components, but also
improve the physics performance of this telescope. 
We also propose detailed feasibility studies towards an ultra-light telescope 
structure, based on computer modeling. 

The successful completion of the proposed studies will give us confidence in
the new technological features, and will allow 
a riskless extrapolation from MAGIC towards ECO-1000, as a final target. A proposal 
for ECO-1000 could be completed as early as 2007, with the goal of having 
an operating instrument in 2009 or 2010.

We further argue strongly in favor of using the existing infrastructure at the
La Palma site for a gradual development towards a multi-telescope European Cherenkov
Observatory. We intend to open this observatory to guest observers, and to make the 
data available to the entire international community of astronomers and astrophysicists.

\section*{Acknowledgement} 
We acknowledge the contribution of Dirk Petry, formerly with the 
MAGIC experiment, and now at University of Maryland, Baltimore County.

\end{document}